\documentclass[journal]{IEEEtran}
\usepackage[noadjust]{cite}

\usepackage{scalerel}
\usepackage{tikz}
\usetikzlibrary{svg.path}
\usepackage{amssymb}

\definecolor{orcidlogocol}{HTML}{A6CE39}
\tikzset{
  orcidlogo/.pic={
    \fill[orcidlogocol] svg{M256,128c0,70.7-57.3,128-128,128C57.3,256,0,198.7,0,128C0,57.3,57.3,0,128,0C198.7,0,256,57.3,256,128z};
    \fill[white] svg{M86.3,186.2H70.9V79.1h15.4v48.4V186.2z}
                 svg{M108.9,79.1h41.6c39.6,0,57,28.3,57,53.6c0,27.5-21.5,53.6-56.8,53.6h-41.8V79.1z M124.3,172.4h24.5c34.9,0,42.9-26.5,42.9-39.7c0-21.5-13.7-39.7-43.7-39.7h-23.7V172.4z}
                 svg{M88.7,56.8c0,5.5-4.5,10.1-10.1,10.1c-5.6,0-10.1-4.6-10.1-10.1c0-5.6,4.5-10.1,10.1-10.1C84.2,46.7,88.7,51.3,88.7,56.8z};
  }
}

\newcommand\orcidicon[1]{\href{https://orcid.org/#1}{\mbox{\scalerel*{
\begin{tikzpicture}[yscale=-1,transform shape]
\pic{orcidlogo};
\end{tikzpicture}
}{|}}}}

\usepackage[hidelinks]{hyperref}

\usepackage{comment}

\usepackage{siunitx}
\usepackage{amsmath}

\usepackage{graphicx}
\usepackage{subcaption}
\usepackage{multirow}

\usepackage{soul}

\usepackage[colorinlistoftodos]{todonotes}

\usepackage{color}
\definecolor{orange}{rgb}{1,0.5,0}
\definecolor{amethyst}{rgb}{0.6,0.4,0.8}
\definecolor{aureolin}{rgb}{0.99,0.93,0.0}
\definecolor{awesome}{rgb}{1.0,0.13,0.32}
\definecolor{ao-green}{rgb}{0.0, 0.5, 0.0}

\begin{document}

\title{Superheterodyne Microwave System for the Detection of Bioparticles with Coplanar Electrodes on a Microfluidic Platform}

\author{C\'esar Palacios \orcidicon{0000-0003-1298-8434}\, \IEEEmembership{Student Member IEEE}, Marc Jofre \orcidicon{0000-0002-8912-6595}\,, Llu\'is Jofre \orcidicon{0000-0003-2437-259X}\,, Jordi Romeu \orcidicon{0000-0003-0197-5961}\, \IEEEmembership{Fellow IEEE} and Luis Jofre-Roca \orcidicon{0000-0002-0547-901X}, \IEEEmembership{Life Fellow IEEE}

\thanks{
C. Palacios, J. Romeu and L. Jofre-Roca are with the Dept. Signal Theory and Communications, Technical University of Catalonia, Barcelona 08034, Spain. 

M. Jofre is with the Dept. Signal Theory and Communications, Technical University of Catalonia, Barcelona 08034; and the Dept. Research and Innovation, Fundaci\'o Privada Hospital Asil de Granollers, Granollers 08402, Spain.

Llu\'is Jofre is with Dept. Fluid Mechanics, Technical University of Catalonia, Barcelona 08019, Spain.

Corresponding author: C. Palacios (email: cesar.augusto.palacios@upc.edu}}


%

\markboth{IEEE Transactions on instrumentation and measurement,~Vol.~-, No.~-, November~2021}%
{Shell \MakeLowercase{\textit{et al.}}: Bare Demo of IEEEtran.cls for IEEE Transactions on Magnetics Journals}
%



\IEEEtitleabstractindextext{%
\begin{abstract}
The combination of microwave and microfluidic technologies has the potential to enable wireless monitoring and interaction with bioparticles, facilitating in this way the exploration of a still largely uncharted territory at the intersection of biology, communication engineering and microscale physics.
Opportunely, the scientific and technical requirements of microfluidics and microwave techniques converge to the need of system miniaturization to achieve the required sensitivity levels. This work, therefore, presents the design and optimization of a measurement system for the detection of bioparticles over the frequency range $0.01$ to \SI{10}{\giga \hertz}, with different coplanar electrodes configurations on a microfluidic platform.
The design of the measurement signal-chain setup is optimized for a novel real-time superheterodyne microwave detection system.
In particular, signal integrity is achieved by means of a microwave-shielded chamber, which is protected from external electromagnetic interference that may potentially impact the coplanar electrodes mounted on the microfluidic device.
Additionally, analytical expressions and experimental validation of the system-level performance are provided and discussed for the different designs of the coplanar electrodes. This technique is applied to measure the electrical field perturbation produced by \SI{10}{\micro \meter} polystyrene beads with a concentration of $10^5$~\si{beads/\milli \liter}, and flowing at a rate of $10$~\si{\micro \liter / \minute}. The achieved SNR is in the order of \SI{40}{\deci \bel} for the three coplanar electrodes considered.
\end{abstract}

}

\IEEEoverridecommandlockouts
\IEEEpubid{\makebox[\columnwidth]{DOI:\href{https://ieeexplore.ieee.org/document/9751686}{10.1109/TIM.2022.3165790}~\copyright2021 IEEE.\hfill}\hspace{\columnsep}\makebox[\columnwidth]{ }}

\maketitle

\IEEEpubidadjcol

\IEEEdisplaynontitleabstractindextext

%
\IEEEpeerreviewmaketitle


\section{Introduction}  \label{sec:introduction}
\IEEEPARstart{T}{he} pairing of microwave sensing and microfluidic platforms enables cutting-edge technological solutions for interfacing with biological systems in medicine, industry, environmental analysis and pharmaceutical applications~\cite{hausser2000hodgkin}. Among the many advantages of interfacing with individual bioparticles provided by the combination of these engineering disciplines, like for example integration of multiple processes into one device and managing small volume of fluids, detection and quantification with high specificity and sensitivity are of significant wide importance~\cite{sun2010single-review}, and are consequently the properties of interest in this work.

Generally, microwave sensors are non-destructive, cost-effective, and highly sensitive instruments that provide rapid sensing capabilities~\cite{s21113759,s21206811}.
Microfluidic devices, on the other hand, offer small sizes, fast speed, automation and high integration characteristics~\cite{mutlu2018oscillatory}.
As a result, these two technologies combined facilitate the development of new microwave-microfluidic systems capable of interacting with microorganisms, providing high throughput, real-time and non-invasive monitoring.
In most standard configurations, the microwave sensing elements mounted in microchannels are composed of parallel facing electrodes~\cite{gawad2004dielectric,C8LC01333K}. However, coplanar electrodes present a number of advantages over traditional parallel facing electrodes~\cite{sun2007analytical}: (i) coplanar electrodes can be integrated into microfluidic chips through standard microfabrication processes; and (ii) their planar structure allows surface scanning by the electrodes to acquire information, and consequently facilitate the possibility of designing wearable sensors in practical applications~\cite{hu2010planar}.
Finally, at the micrometric scale, focusing and sorting of bioparticles is a challenge that can be overcome through the advantages offered by microfluidic technologies, which typically make use of microchannels with cross-sections in the order of \SI{100}{\micro \meter} or smaller.
These characteristics enable automated control and position of bioparticles without mechanical moving parts, and presenting little noise generation in the process~\cite{jofre2021wireless}.

Different examples of electromagnetic bioparticle detection methods that are, at present, operational have been described in the literature~\cite{alahnomi2021review,qian2014dielectrophoresis,nikolic2009microwave,ermolina2005electrokinetic,brown2002remote}.
In general, the most common approach is based on optical methods consisting in illuminating the microchannel with light at frequencies typically within the visible range of the electromagnetic spectrum, and sensing in the far-field regime the optical phenomena caused by the bioparticles as they flow along the centerline of the microchannel~\cite{holzner2017elasto,etcheverry2017high,matsumoto2016microfluidic,Perez:17}.
Electrical-based strategies have also been proposed, which perform measurements at low electromagnetic frequencies (typically $f < \SI{0.1}{\mega \hertz}$, with $f$ the frequency), and present wavelengths much larger than the dimensions of the system.
In this type of strategies, the principal contribution in the detection process is provided by the electromagnetic dielectric properties of the particles~\cite{haandbaek2014resonance,holmes2010single,spencer2020high}.
On the other hand, microwave methods typically measure field disturbances, in the near-field region, at frequencies in the range $\SI{0.1}{\giga \hertz} < f < \SI{100}{\giga \hertz}$, where the wavelength is comparable to the dimensions of the microfluidic channel and sensing elements (e.g., electrodes)~\cite{narang2018sensitive,li2018differentiation,jankovic2017microwave}.
Although there is no specific technique that reins in all domains, over the past decades, many advances have been achieved by utilizing optical and low/medium-frequency electrical techniques. In general, optical approaches face technological difficulties posed by the complexity of arranging the components required, while electrical techniques present drawbacks related to selectivity, stability and reproducibility.
As an alternative, or in combination with the latter, microwaves can potentially overcome some of these limitations~\cite{jofre2021wireless}.
In this regard, this work is focused on microwave techniques at the microfluidic scale, which is of interest for particle detection applications, and especially relevant for sensing and interacting with particles of size comparable to that of microorganisms, e.g., E. Coli, referred to in this paper as bioparticles.

The objectives of this work, therefore, are to (i) derive a microwave-based theoretical framework for designing and assessing the performance of coplanar electrodes configurations in microfluidic platforms to detect bioparticles, (ii) provide an experimental demonstration of the capabilities achieved in combination with a proposed microwave superheterodyne receiver, and (iii) discuss the results at microwave frequencies.
In this regard, the paper is organized as follows.
First, analytical expressions of the spatial electrical field distributions of three different sensing electrodes to detect particles are provided in Section~\ref{sec:theory}.
Next, Section~\ref{sec:system} presents the microfluidic particle-focusing platform and describes the microwave detection system.
Experimental demonstration of the system and presentation and discussion of the results are, respectively, provided in Sections~\ref{sec:results} and~\ref{sec:discussion}.
Finally, Section~\ref{sec:conclusions} summarizes the work and provides conclusions.

\section{Theoretical description}     \label{sec:theory}
\subsection{Electric field sensing}
The sensing technique utilized in this work is based on the capacitance variations measured from the perturbation that a single bioparticle produces in the electric field, which is mostly concentrated in the near-field as described below.
In general, the Near Field Region (NFR) and the Far Field Region (FFR) are defined as $\textrm{NFR}<2D^2/\lambda<\textrm{FFR}$, where $D$ is the maximum physical linear dimension of the electrical configuration, and $\lambda$ is the wavelength in the medium.
The NFR is divided into two parts: (i) the Reactive Near Field, where the electric and magnetic fields are $90^o$ out of phase; and (ii) the Radiative Near Field, where the electromagnetic field starts the transition from reactive to radiative.
The limit between these two regions is typically defined as $0.62\sqrt{D^3/\lambda}$.
The microconfined flows considered in this work reduce the distances at which the bioparticles are found with respect to the coplanar electrodes; viz. in the order of \SI{40}{\micro \meter}.
As a result, instead of studying the far-field region~\cite{simeoni2010equilateral}, the analysis of the electric fields is carried out for the near-field region.
Moreover, while the electric field distribution is mostly uniform for parallel facing electrodes, the electric field distribution varies spatially as a squared function of distance to the electrodes in coplanar configurations~\cite{jofre2021wireless}, resulting in a nonlinear response that can be leveraged for sensing applications.

Considering a microfluidic channel filled with a solution (e.g., water) of dielectric complex relative permittivity $\varepsilon^{*}_{r,m}$, the initial stored energy of the region can be written as $U_0=-1/2C_0V_0^2$, where $V_0$ is the applied voltage and $C_0$ is the capacitance of the electrodes. Therefore, when a particle with dielectric complex relative permittivity $\varepsilon^{*}_{r,p}$ passes through the detection region, it causes a capacitance variation $\Delta C$, and the energy variation becomes $\Delta U=-1/2\Delta C V_0^2$.
In particular, the applied electric field produces an effective induced dipolar moment $\vec{p} = \alpha \vec{E}$~\cite{griffiths2005introduction}, where $\alpha$ is the polarizability, which for a spherical particle of radius $a$ is equal to $\alpha =4\pi \varepsilon_0 \varepsilon^{*}_{r,p} a^3$, where $\varepsilon_0$ is the permittivity of vacuum.
Consequently, the perturbed energy can be computed as $\Delta U = -1/2\alpha E_0 (\vec{r})^2$, where $E_0 (\vec{r})$ is the electric field intensity in the medium.
The induced change in capacitance is then found to be
\begin{equation}
\Delta C = 4 \pi  a^3 \Re \left \{\varepsilon^{*}_{r,m}  K_{CM}(\omega)\right \} \left |E_0  \right |^2 /V_0^2,
\label{eq1:var_cap}
\end{equation}
where $K_{CM}(\omega) =[\varepsilon^{*}_{r,p}(\omega)-\varepsilon^{*}_{r,m}(\omega)]/[\varepsilon^{*}_{r,p}(\omega)+2\varepsilon^{*}_{r,m}(\omega)]$ is the complex Clausius-Mossotti factor describing the particle-medium dielectric contrast, and $\omega$ is the angular frequency.

\begin{figure}[t!]
     \centering
     \begin{subfigure}[b]{0.45\columnwidth}
         \centering
         \includegraphics[width=\columnwidth]{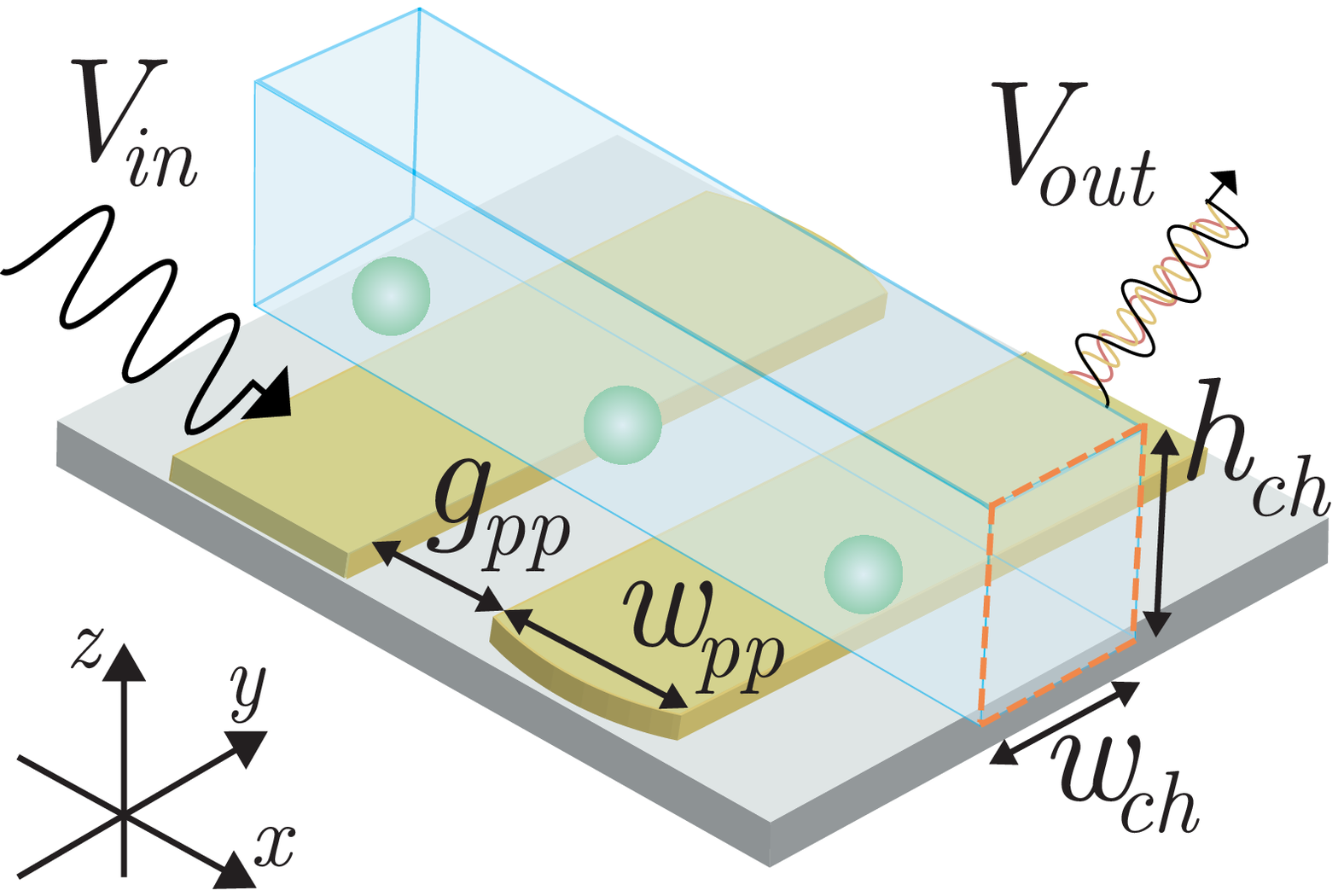}
         \caption{}
         \label{fig:PP_layout}
     \end{subfigure}
     \hfill
     \begin{subfigure}[b]{0.5\columnwidth}
         \centering
         \includegraphics[width=\columnwidth]{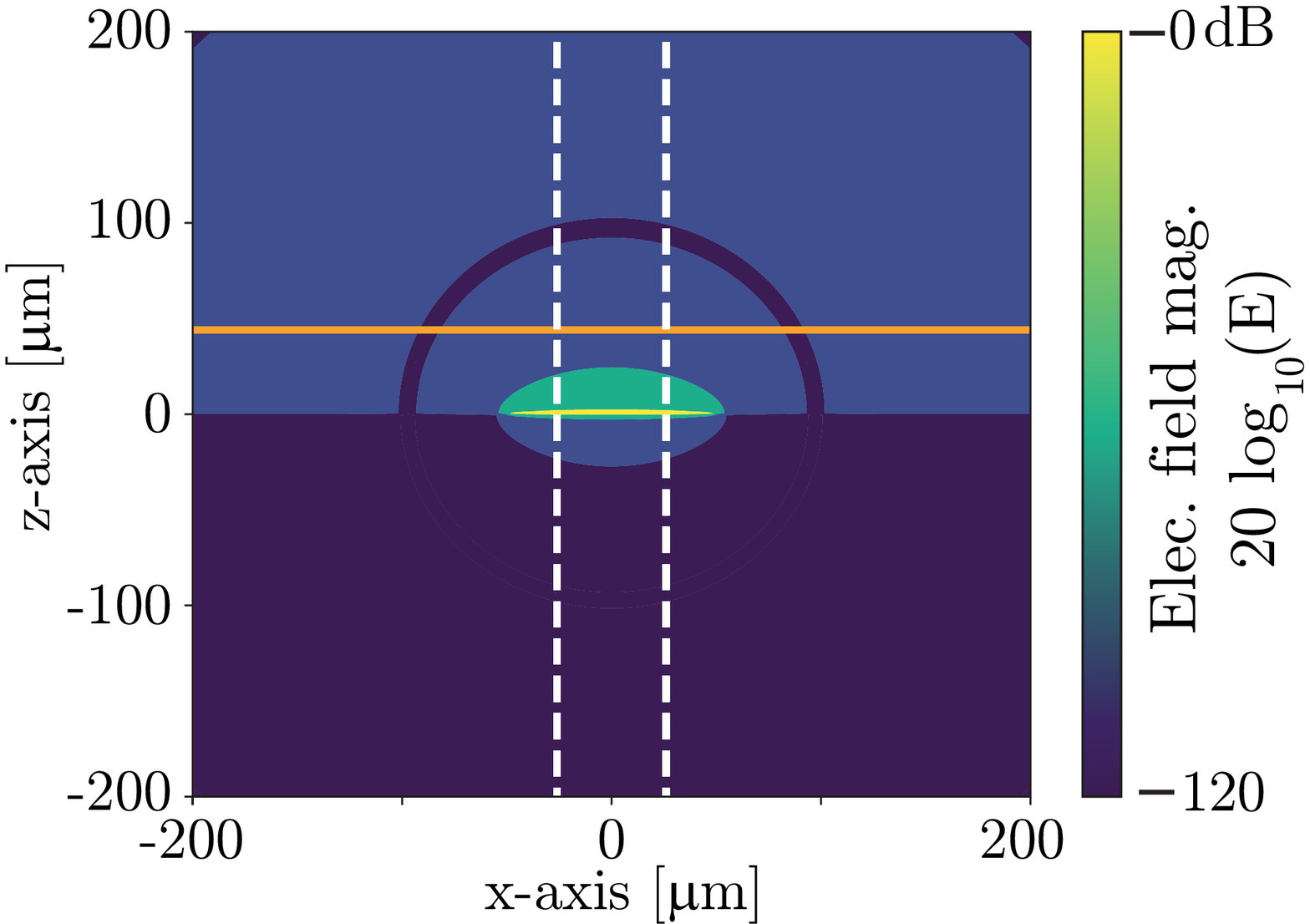}
         \caption{}
         \label{fig:PP_EField}
     \end{subfigure}
     \hfill
     \begin{subfigure}[b]{0.45\columnwidth}
         \centering
         \includegraphics[width=\columnwidth]{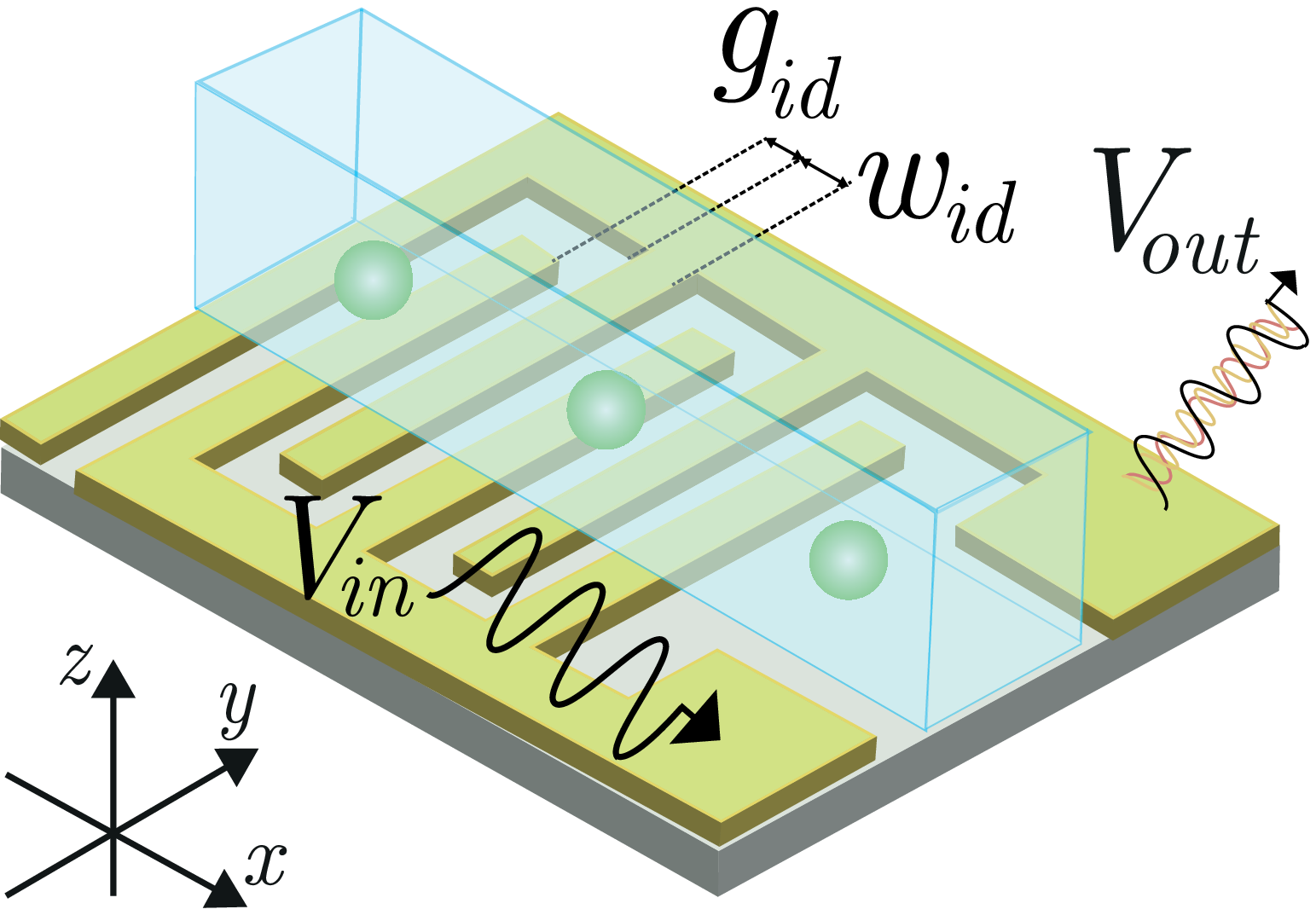}
         \caption{}
         \label{fig:IDE_layout}
     \end{subfigure}
     \hfill
     \begin{subfigure}[b]{0.5\columnwidth}
         \centering
         \includegraphics[width=\columnwidth]{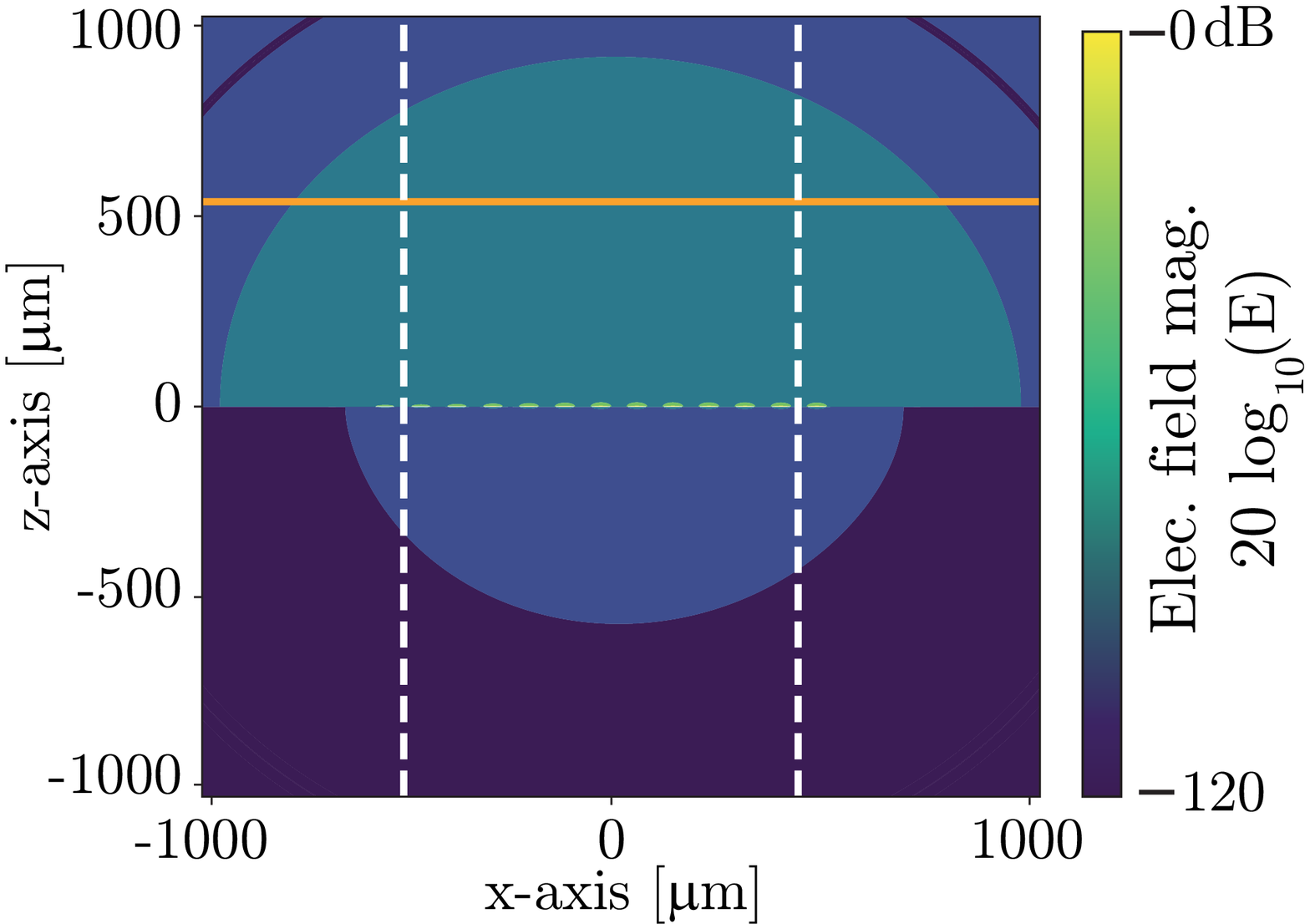}
         \caption{}
         \label{fig:IDE_EField}
     \end{subfigure}
     \hfill
     \begin{subfigure}[b]{0.45\columnwidth}
         \centering
         \includegraphics[width=\columnwidth]{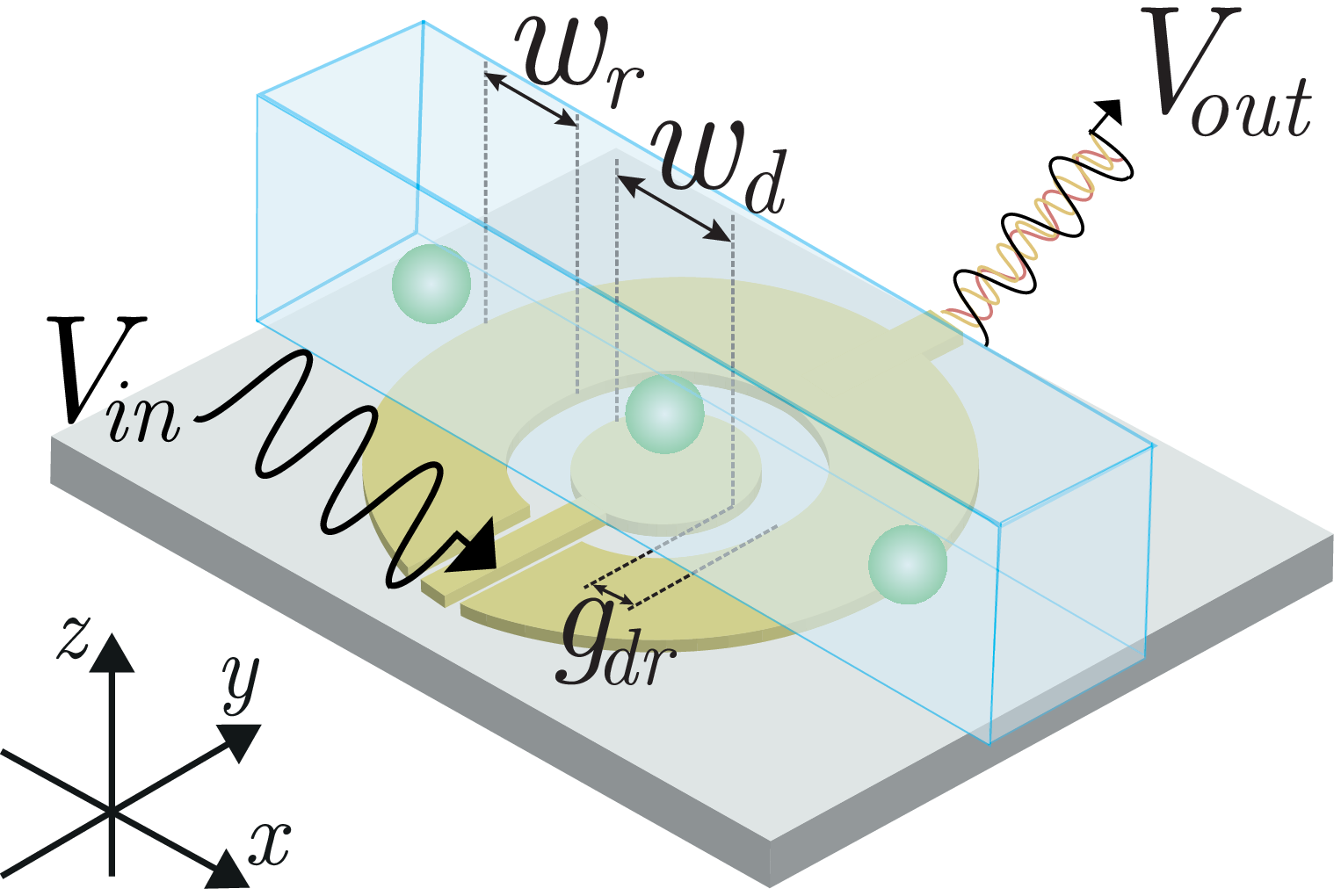}
         \caption{}
         \label{fig:DR_layout}
     \end{subfigure}
     \hfill
     \begin{subfigure}[b]{0.5\columnwidth}
         \centering
         \includegraphics[width=\columnwidth]{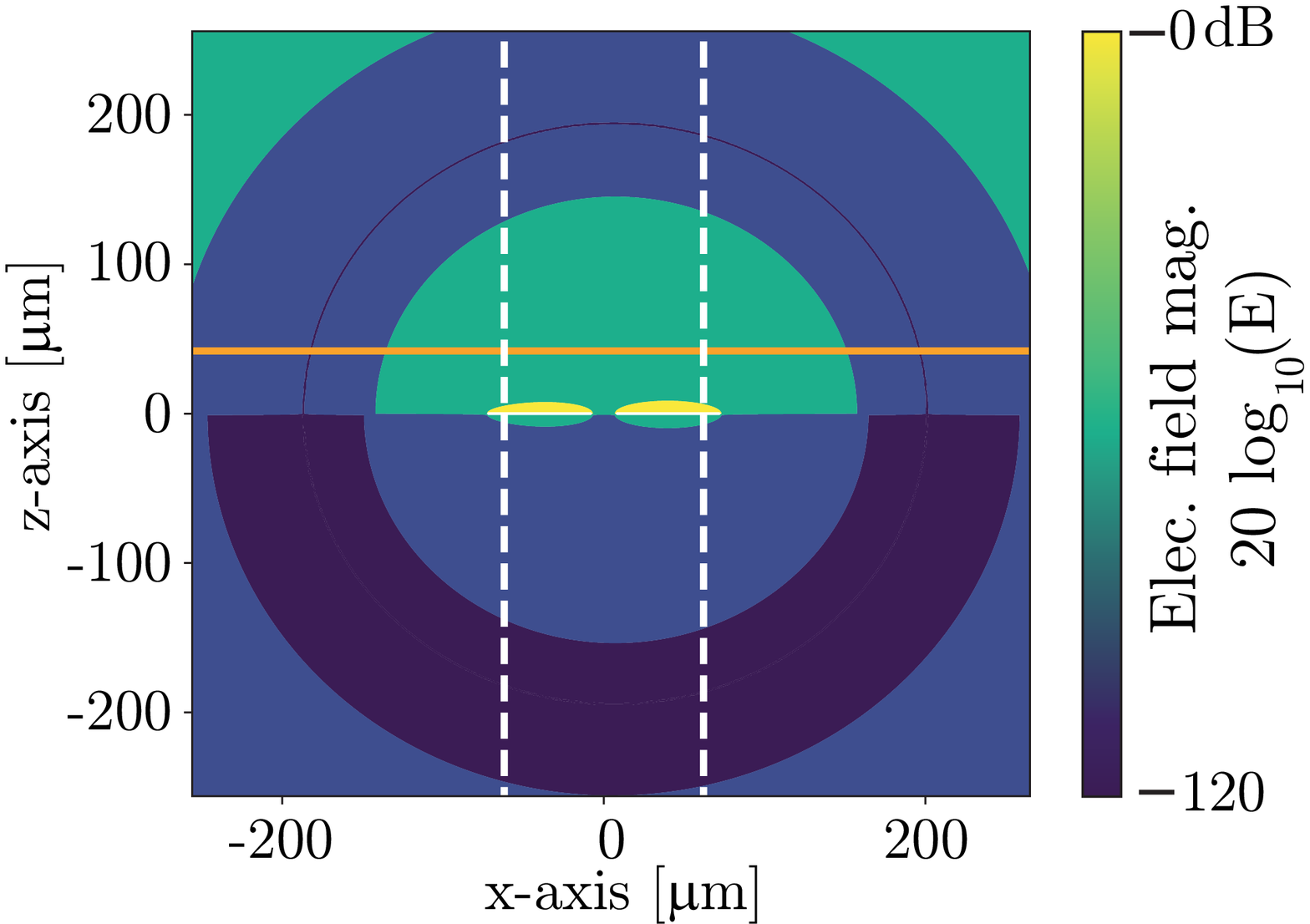}
         \caption{}
         \label{fig:DR_EField}
     \end{subfigure}
    \caption{Geometrical configurations (a, c, e) and electric field distributions and hotspots (b, d, f) at \SI{1}{\giga \hertz} of the PP (a, b), ID (c, d), and DR (e, f) electrodes mounted on the microfluidic channel. Orange and white lines represent, respectively, the maximum penetration depth and width of the hotspot.}
    \label{fig:Electrodes_EDistribution}
\end{figure}

To detect these particles, three different configurations of coplanar electrodes, with $\sim$\SI{100}{\micro \meter} overall width and \SI{10}-\SI{75}{\micro \meter} gap separation, mounted on the bottom wall of a microfluidic channel are studied.
The first coplanar configuration, named Parallel Plates (PP), consists of two parallel electrodes of width $w_{pp}$ coplanarly positioned with a gap separation $g_{pp}$ between them as depicted in Fig.~\ref{fig:PP_layout}.
The second electrode configuration, referred to as Interdigitated (ID), is composed of multiple sets of parallel coplanar electrodes, each pair having a width $w_{id}$ and a gap separation $g_{id}$~\cite{starzyk2008parametrisation,mamishev2002uncertainty}.
As shown in Fig.~\ref{fig:IDE_layout}, two main connections feed the elements of each pair.
The ID configuration analyzed is characterized by $w_{id}=g_{id}$.
The setup displacement from the origin, as shown in Fig.~\ref{fig:IDE_EField}, forms an elliptical electrical field distribution between the electrodes within the microfluidic channel.
The penetration depth of the resulting hotspot varies with $w_{id}$ and $g_{id}$.
In this configuration, almost all energy is concentrated within $2(w_{id}+g_{id})$, making ID electrodes significantly efficient for measurements of small-volume samples~\cite{claudel2020interdigitated}.
The third configuration considered, as shown in Fig.~\ref{fig:DR_layout} and named Disk Ring (DR), presents the inherent advantage of rotational symmetry and a larger sensing area~\cite{chen2010analysis}.
DR electrodes are formed by an inner central disk of diameter $w_d$, and an outer ring of width $w_r$, although for this analysis it is considered $w_d$=$w_r$=$w_{dr}$; the gap between the disk and the ring is $g_{dr}$ as depicted in Fig.~\ref{fig:DR_layout}.
The field lines go from the central disk to the outer ring, as shown in Fig. \ref{fig:DR_EField}.
This particular characteristic warrants analyzing this type of sensors as an open-ended coaxial probe~\cite{skierucha2004comparison}, where the capacitance value describes the geometry of the probe.
Typically, the larger the test dimensions (comparable to the size of bioparticles), the larger the capacitance value, and as a result more accurate measurements of small variations of real relative permittivity can be obtained.

\subsection{Retardation effects}
In this section, the electromagnetic retardation effects of the fields produced by the electrodes are presented.
The derivation of the radiation field resulting from time-varying charges and currents is based on Green's functions to describe the concept of retarded vector potential.
Starting from Maxwell's equations and following the derivations of~\cite{ulaby2005electromagnetics}, the scalar $\phi(\vec{r},t)$ and vector $A(\vec{r},t)$ potentials of a charge $\rho(\vec{r},t)$ and current $\vec{J}(\vec{r},t)$ varying in time correspond, respectively, to
\begin{equation}
   \phi(\vec{r},t)=\int_{\vec{r'}}\rho (\vec{r'},t-| \vec{r}-\vec{r'} |/c )/ ( 4\pi | \vec{r}-\vec{r'} | )d^3\vec{r'}
    \label{eq:retarded}
\end{equation}
and
\begin{equation}
   A(\vec{r},t)=\int_{\vec{r'}}\vec{J} (\vec{r'},t- | \vec{r}-\vec{r'} |/c )/ ( 4\pi | \vec{r}-\vec{r'} | ) d^3\vec{r'}.
    \label{eq:retarded2}
\end{equation}

The retarded potential is the effect observed at position $r$ at time $t$ due to the disturbance originated at a retarded time $t'=t-\left | r-r' \right |/c$ and location $r'$ propagating at speed $c$.
The difference $t-t'$ can be explained as the time that the wave takes to propagate from the source to the point where it is measured.
The size of the electrodes studied is much shorter than the wavelength of the electromagnetic wave of the applied signal.
As a result, the electrodes are assumed to be represented as short dipole antennas~\cite{jackson1999classical}.
In this regard, the electric field as a function of position $\vec{r}$ and time $t$ for a short dipole antenna is modeled as
\begin{align}
 \vec{E}(\vec{r},t)&=\Re \left \{j c \frac{\mu_0 k I_0 L}{4 \pi r} \sin\theta e^{-jkr}e^{j \omega t} \right \} \hat{\theta}, \label{eq:retarded_Efield}
\end{align}
where $\mu_0$ is the vacuum permeability, $k$ is the wavenumber in the medium, $I_0$ is the time harmonic current, $L$ is the length of the equivalent electrical dipole, and $\theta$ is the angle between $x$-axis and $\vec{r}$ in the $x-z$ plane.

It is important to note, based on Eq.~\ref{eq:retarded_Efield}, that wave propagation is function of $e^{-jkr}$ and $e^{j \omega t}$.
In this regard, considering that the time that a particle remains within the electric field radiated by the electrodes is very large compared to the time period of the source signal, the temporal effect of the factor $e^{j \omega t}$ is negligible. However, the factor $e^{-jkr}$ modifies the electric field distribution depending on the wavelength. Therefore, if the real part of Eq.~\ref{eq:retarded_Efield} is considered, the electric field changes with $\cos(2\pi r/\lambda)$, where $\lambda$ is the wavelength in the medium of the applied signal. The retardation effect will need to be considered when $r/\lambda > 0.01$, that may be obtained for dimensions of the electrode area of millimeters and operation frequencies of GHz. Consequently, the retardation effect only becomes negligible in the limit when the electrodes are much smaller than the wavelength in the medium, i.e., $2\pi r/\lambda \ll 1$.

\subsection{Electric field distributions}
The $E_z$ electric field distribution in the physical plane $x-z$, corresponding to a pair of coplanar electrodes of width $w$ and a gap separation $g$, can be obtained using the technique of conformal transformation~\cite{linderholm2005comment,hong2005ac} as
\begin{equation}
    E_z=E_W \frac{dW(Z)}{dZ}, \label{eq:ConTransform}
\end{equation}
where $E_W$ is the electric field in the transformed complex $W$-plane; i.e., $W=u(x,y)+j v(x,y)$ with $u$ the electrical potential function and $v$ proportional to the electrical flux function. In the $W$-plane, $E_W$ is defined as
\begin{equation}
    E_W=\frac{V_{0}}{2K(q^2)},
\end{equation}
where $q=g/(g+2w)$ is related to the electrode layout, $V_{0}$ is the applied voltage, and $K(q^2)$ is the complete elliptic integral of the first kind applied to $q^2$.
Hence, the electric field distribution over each of the three configurations depicted in Fig.~\ref{fig:Electrodes_EDistribution} can be defined in terms of short dipoles, considering the retardation effects and the conformal transformation as explained below.

The description of the PP electrode configuration in the $W$-plane~\cite{chen2004capacitive} is defined as
\begin{equation}
    W_{PP}=V_0 - \frac{2V_{0}}{\pi} \cos^{-1}\left ( \frac{Z}{g_{pp}/2}\right ),
\end{equation}
where $Z=X+j Y$ is the coordinate position in the complex $W$-plane.
Using Eq.~\ref{eq:ConTransform} and considering the two electrodes as a short dipole antenna, the E-field distribution of the PP configuration, as depicted in Fig.~\ref{fig:PP_EField}, can be described as
\begin{equation}
    E_{PP}=\frac{V_{0}}{2K(q_{pp}^2)}\frac{1}{\sqrt{(1-l_{pp}^2) (1-q_{pp}^2 l_{pp}^2)}}\cos\left [ \frac{2\pi \sqrt{\varepsilon^{'}_{m}}}{\lambda_0} r\right ],
\end{equation}
where $\lambda_0$ is the wavelength in free space, $\varepsilon^{'}_{m}$ is the dielectric real permittivity of the medium, and variables $q_{pp}$, $l_{pp}$ and $r$ are defined as $q_{pp}=g_{pp}/(g_{pp}+2w_{pp})$, $l_{pp}=1/2-1/\pi \cos^{-1}[2 Z/g_{pp}]$ and $r=\sqrt{X^2+Y^2}$, respectively.

The ID configuration is the sum of several parallel electrodes, displaced from the origin, with a conformal transformation described as
\begin{equation}
    W_{ID}=V_0 - \sum_n \frac{2V_{0}}{\pi} \cos^{-1}\left (\frac{X_{id,n}+jY}{g_{id}/2}\right ),
\end{equation}
where $n=\left [ -N,\dots,N \right ] \in \mathbb{N}$ with $N+1$ even-odd pairs, and $X_{id,n}$ correspond to the symmetrical location of each electrode from the origin of coordinates as depicted in Fig.~\ref{fig:IDE_EField}. In particular, the symmetrical locations are defined as: (i) $X_{id,n}=X-n \left ( w_{id}+g_{id} \right )$ for even $n \geq 0$, (ii) $X_{id,n}=X-(n+1) \left ( w_{id}+g_{id} \right )$ for even $n<0$, (iii) $X_{id,n}=-X-(n-1) \left ( w_{id}+g_{id} \right )$ for odd $n > 0$, and (iv) $X_{id,n}=-X-n \left ( w_{id}+g_{id} \right )$ for odd $n<0$.
The corresponding E-field distribution is computed as
\begin{align}
    E_{ID}=&\frac{V_{0}}{2K(q_{id}^2)}\sum_n \frac{1}{\sqrt{(1-l_{id,n}^2) (1-q_{id}^2l_{id,n}^2)}}\\
    &\cos\left [\frac{2\pi \sqrt{\varepsilon{'}_{m}}}{\lambda_0} \left (r - n \left [ w_{id}+g_{id} \right ] \right ) \right ], \nonumber
\end{align}
where variables $q_{id}$ and $l_{id,n}$ are defined as $q_{id}=g_{id}/(g_{id}+2w_{id})$ and $l_{id,n}=1/2-1/\pi \cos^{-1}[2 (X_{id,n}+jY)/g_{id}]$.

Finally, the DR configuration consists in a central electrode which is coupled to two side electrodes.
This setup can be approached as two pairs of PP displaced from the origin and oriented toward the center.
In this case, the conformal transformation corresponds to
\begin{equation}
    W_{DR}=V_0 - \sum_n \frac{2V_{0}}{\pi} \cos^{-1}\left ( \frac{X_{dr,n}+jY}{g_{dr}/2}\right ),
\end{equation}
where $n=\left [ -1,1 \right ]$, $X_{dr,-1}=-X - \left ( w_{dr}+g_{dr} \right )/2$ and $X_{dr,1}=X - \left ( w_{dr}+g_{dr} \right )/2$, which are utilized for the equivalent two pairs of electrodes schematically represented in Fig.~\ref{fig:DR_EField}.
Similarly, the E-field distribution for the DR configuration is calculated as
\begin{align}
    E_{DR}=&\frac{V_{0}}{2K(q_{dr}^2)}\sum_n \frac{1}{\sqrt{(1-l_{dr,n}^2) (1-q_{dr}^2 l_{dr,n}^2)}}\\
    &\cos\left [ \frac{2\pi \sqrt{\varepsilon^{'}_{m}}}{\lambda_0} \left ( r- n \frac{w_{dr}+g_{dr}}{2}\right )\right], \nonumber
\end{align}
where $q_{dr}=g_{dr}/(g_{dr}+2w_{dr})$ and $l_{dr,n}=1/2-1/\pi \cos^{-1}[2 (X_{dr,n}+jY)/g_{dr}]$.

The analytic expressions above are valid for the upper spatial hemisphere considering an infinite channel with a dielectric real permittivity of the medium $\varepsilon^{'}_{m}$, and are utilized in Section~\ref{sec:results} to analyze the experimental data.
In particular, these expressions are utilized to quantitatively assess and discuss, in terms of strength of signal variations produced by the sensed bioparticles, the experimental results obtained for the three configurations of coplanar electrodes considered.

\section{System operation}\label{sec:system}
\begin{figure}[!t]
     \centering
     \begin{subfigure}[b]{1\columnwidth}
         \centering
         \includegraphics[width=\columnwidth]{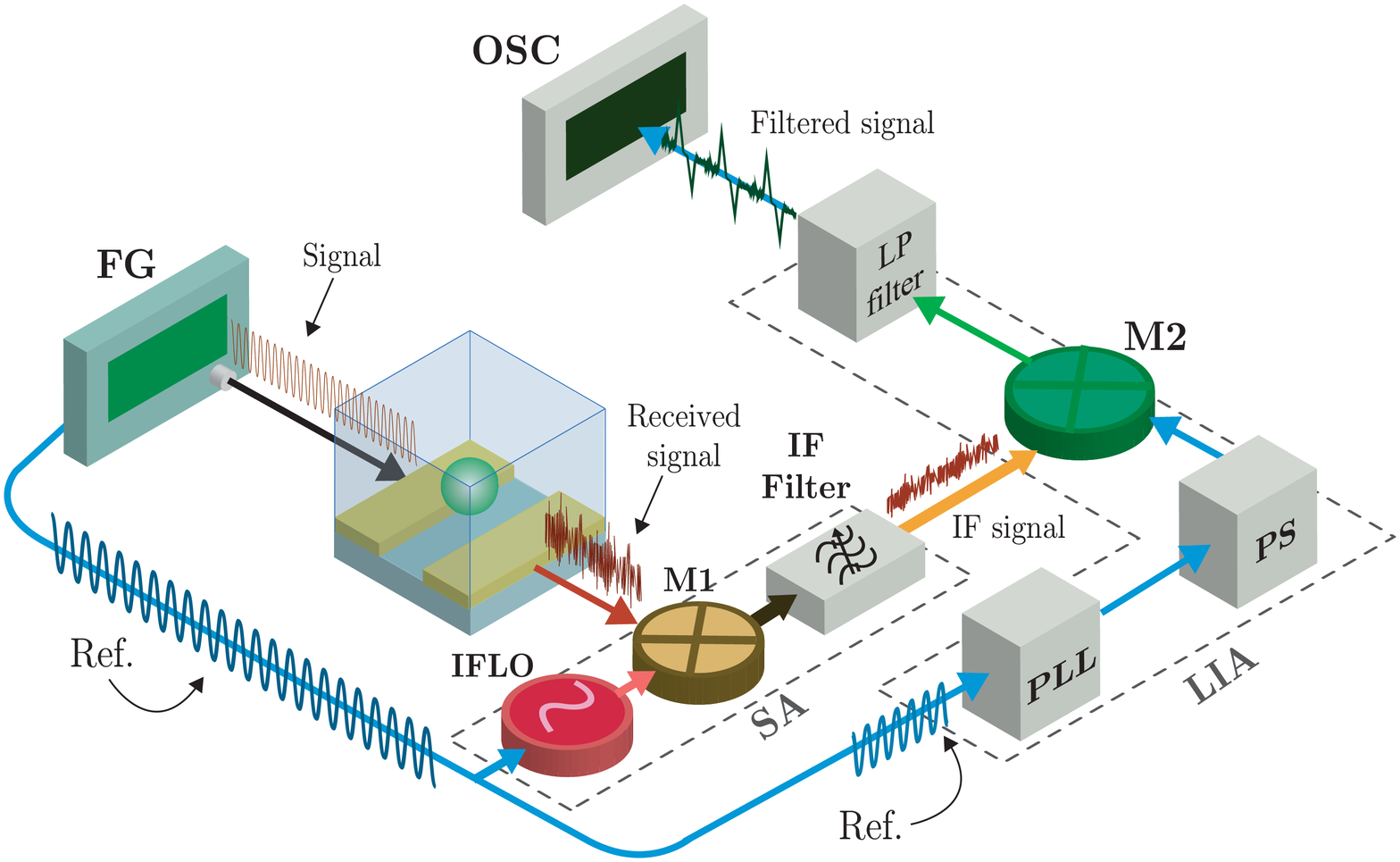}
         \caption{}
         \label{fig:schematic}
     \end{subfigure}
     \hfill
    \begin{subfigure}[b]{0.55\columnwidth}
         \centering
         \includegraphics[width=\columnwidth]{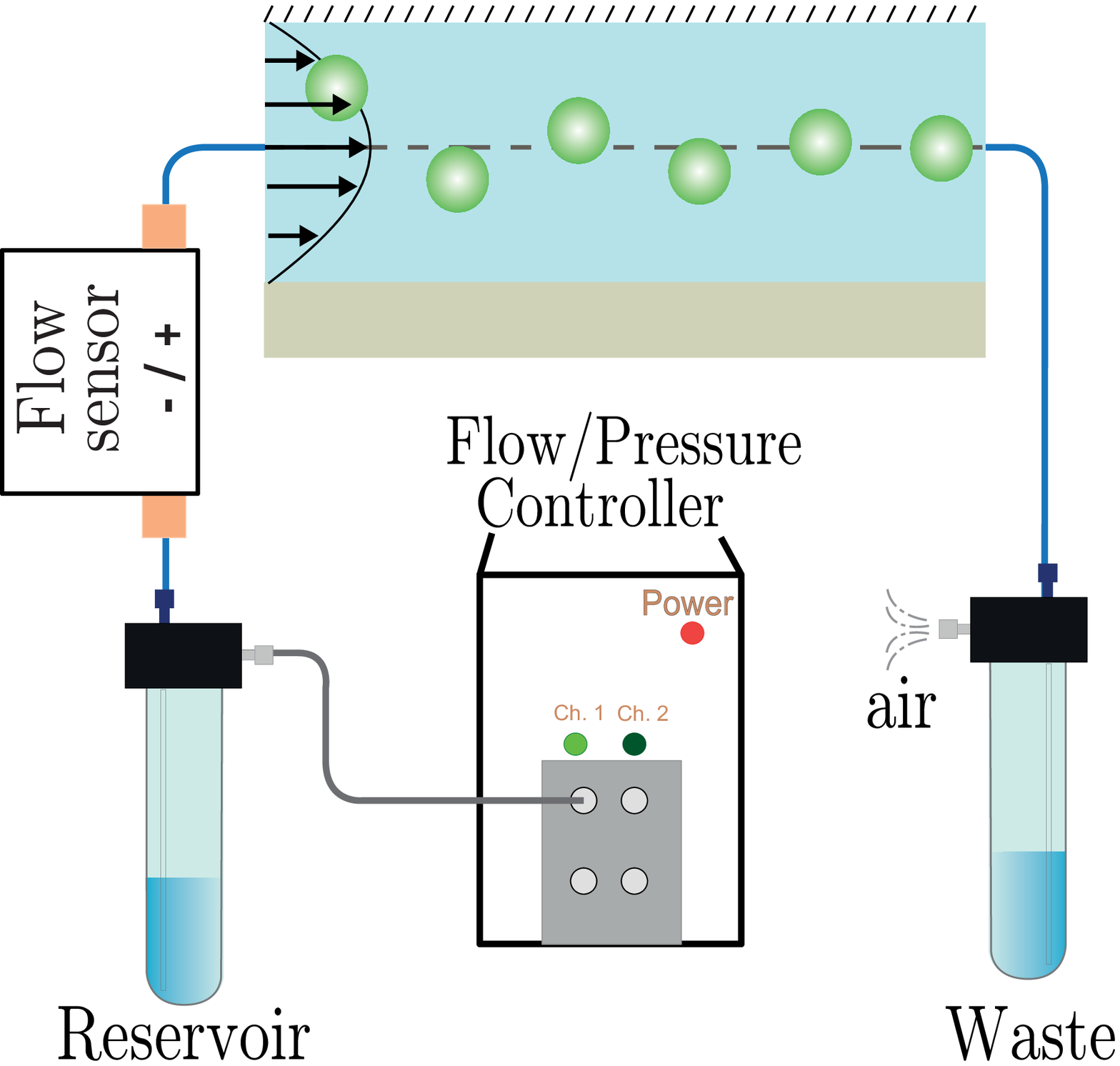}
         \caption{}
         \label{fig:microfluidics}
    \end{subfigure}
    \hfill
    \begin{subfigure}[b]{0.43\columnwidth}
         \centering
         \includegraphics[width=\columnwidth]{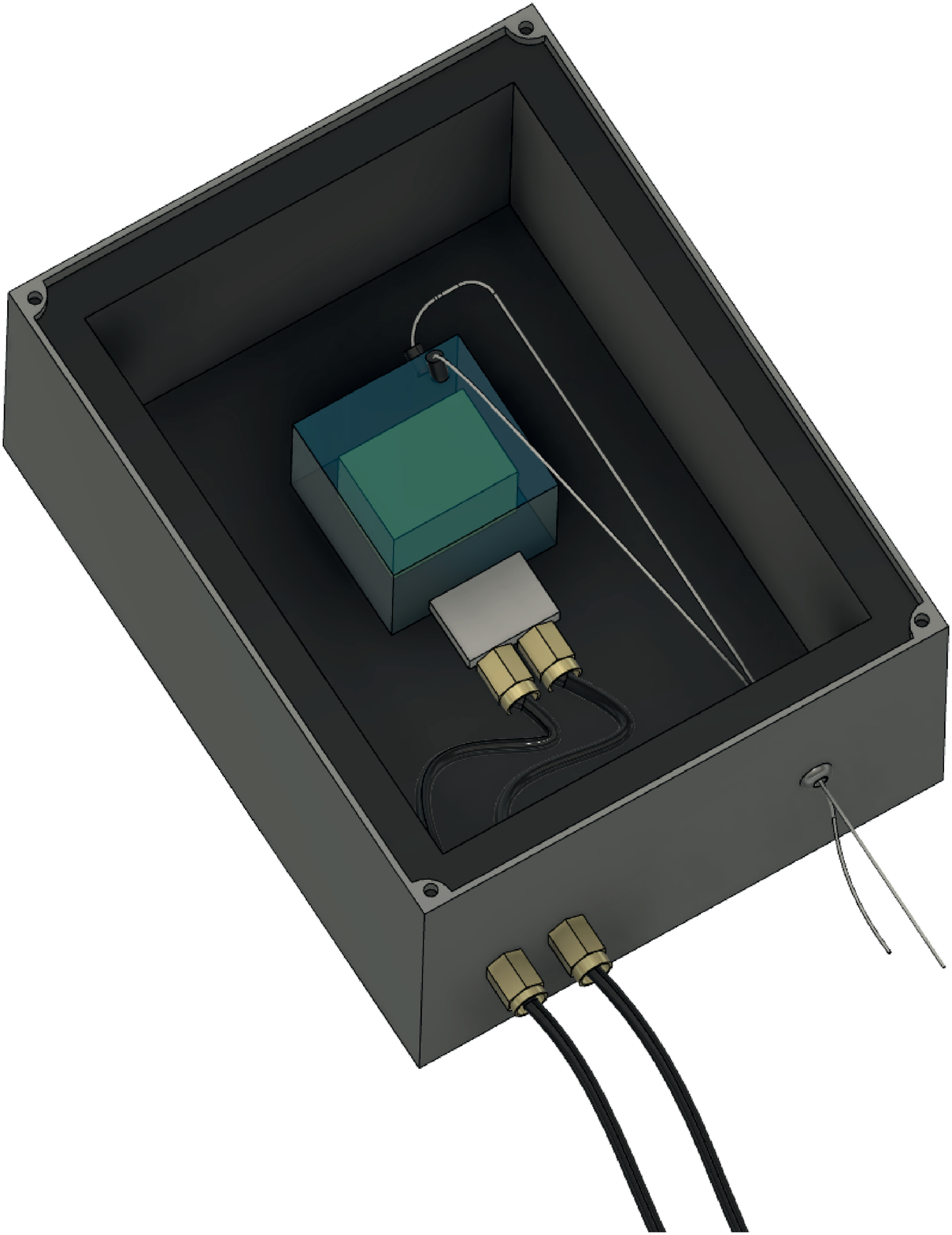}
         \caption{}
         \label{fig:enclosure}
    \end{subfigure}
    \caption{Schematics of the superheterodyne microwave system (a) on a microfluidic platform (b) with coplanar electrodes and shielding enclosure (c).}
    \label{fig:experimental_setup}
\end{figure}

The experimental rig developed is schematically illustrated in Fig.~\ref{fig:experimental_setup}, in which the different configurations of the coplanar electrodes presented in Section~\ref{sec:theory} have been mounted for their characterization.
In particular, the system consists mainly of (i) microwave instrumentation for generating and detecting signals, and (ii) a flow-controlled microfluidic pump station combined with a sealed microchip holder.
Detailed information about the characteristics of these two main subsystems is provided in the paragraphs below.

\subsection{Microwave architecture}
The microwave system is based on an original superheterodyne receiver architecture that allows to baseband-detect signals in the GHz range, overcoming in this way conventional limitations of homodyne systems for detecting high-frequency microwave signals. It is composed of a microwave function generator (FG) (SMB 100A, Rohde \& Schwarz, Germany), a spectrum analyzer (SA) (FSW43, Rohde \& Schwarz, Germany), a lock-in amplifier (LIA) (SR844, Stanford Research Systems, USA), and an oscilloscope (OSC) (RTC1022, Rohde \& Schwarz, Germany) for visualizing and measuring the received signals. The different components have been connected by using \SI{50}{\ohm}-matched coaxial cables, and the master reference \SI{10}{\mega \hertz} clock from the FG has been propagated along the subsystems to ensure proper synchronization of the equipment.
The FG provides \SI{50}{\ohm}-matched sinusoidal signals up to \SI{20}{\giga \hertz} with a maximum power of $13$ dBm.
The SA provides a working frequency from \SI{10}{\mega \hertz} to \SI{43.5}{\giga \hertz} with a receiver dynamic range of \SI{100}{\deci \bel}, and the LIA can operate from \SI{25}{\kilo \hertz} to \SI{200}{\mega \hertz} with a baseline white-noise sensitivity limited to $2\sim\SI{8}{\nano V/\sqrt{\hertz}}$. On top of the microfluidic chip, a \SI{10}{\giga \hertz} resonance free grounded coplanar waveguide has been used to interface the electrodes with the input/output coaxial cables. Overall, the system allows a proper operation up to the foreseen frequency of \SI{10}{\giga \hertz}.

The microwave signal from the FG is fed into the coplanar electrodes. Simultaneously in time, the \SI{10}{\mega \hertz} reference clock of the FG is used to synchronize the SA. The SA is configured to work in zero span mode, and converts the received signal with an intermediate frequency local oscillator (IFLO) to a \SI{10}{\mega \hertz} signal, with \SI{3}{\kilo \hertz} of bandwidth, by means of the multiplier (M1). The M1 output signal is a midband replica of the measured signal. In addition, the SA \SI{10}{\mega \hertz} synchronized reference clock signal is utilized as the input reference for the LIA, which includes a phase-locked loop (PLL), a phase-shifter (PS), a multiplier (M2), and an adjustable low-pass filter producing a baseband signal at the output. Finally, an oscilloscope is used to visualize and digitize the measured signals in magnitude and phase as reported in Fig.~\ref{fig:experimental_setup}.
All the electromagnetic systems are properly shielded with a coated grounded enclosure that reduces interference in \SI{40}{\deci \bel}; a 3D design of the shielding is shown in Fig.~\ref{fig:enclosure}.
This shielding avoids electromagnetic interferences present in the laboratory environment to interact with the system, and reduces the impact of potential picked-up noise resulting from the utilization of exposed electrical elements.

\subsection{Microfluidic platform}
The microfluidic part of the experiment is based on a pressure-driven flow-controlled pump, which is capable of providing \SI{30}{psi} of maximum pressure to drive the fluid through the microchannel; a schematic of the system is presented in Fig.~\ref{fig:microfluidics}.
The different microfluidic components are connected using polyetheretherketone (PEEK) and polytetrafluoroethylene (PTFE) tubing of 1/32” of internal diameter; the internal diameter of the tubes is constant throughout the system to reduce the quantity of bubbles generated and enhance the stability of the flow. The hydraulic diameter of the microchannels is \SI{40}{\micro \meter}.

As a surrogate for bioparticles, the experiments were performed utilizing polystyrene beads with average diameter of \SI{10}{\micro \meter} and concentration of $10^5$~\si{beads/\milli \liter}.
The microparticles were suspended in a medium that emulates the functional conditions of living organisms.
In a set of initial experiments, air bubbles were introduced into the tubes to provide a higher permittivity contrast to calibrate the system.
The preparation of the liquid medium followed the method described in a previous work~\cite{jofre2021wireless}, which is composed of \SI{0.2}{\micro \meter}-filtered water with $500$ ppm of polyethylene oxide (PEO) cured for more than 4 weeks; the PEO was taken from a $10^4$~ppm of $0.4$~MDa prepared stock.

The procedure to run an experiment starts by pouring the mixture solution prepared into a reservoir of \SI{10}{\milli \liter}.
Next, the mixture is introduced into the fluid tubes from this reservoir at a flow rate of \SI{10}{\micro\liter/\min} by the pressure pump, which is connected to a flow controller with a resolution of \SI{1}{\micro \liter/\min}.
Then, the mixture flows through the microchannel in which a balance of hydrodynamic forces constraints the microparticles to focus at its centerline~\cite{jofre2021wireless}, and passing through the electromagnetic fields generated by the electrodes.
Finally, the mixture is collected in a second reservoir (waste) at the outlet of the microchannel to be later recycled.

\section{Experimental results}    \label{sec:results}
\subsection{Response characterization of the electrodes}
\begin{figure}[!t]
     \centering
     \begin{subfigure}[b]{0.85\columnwidth}
         \centering
         \includegraphics[width=\columnwidth]{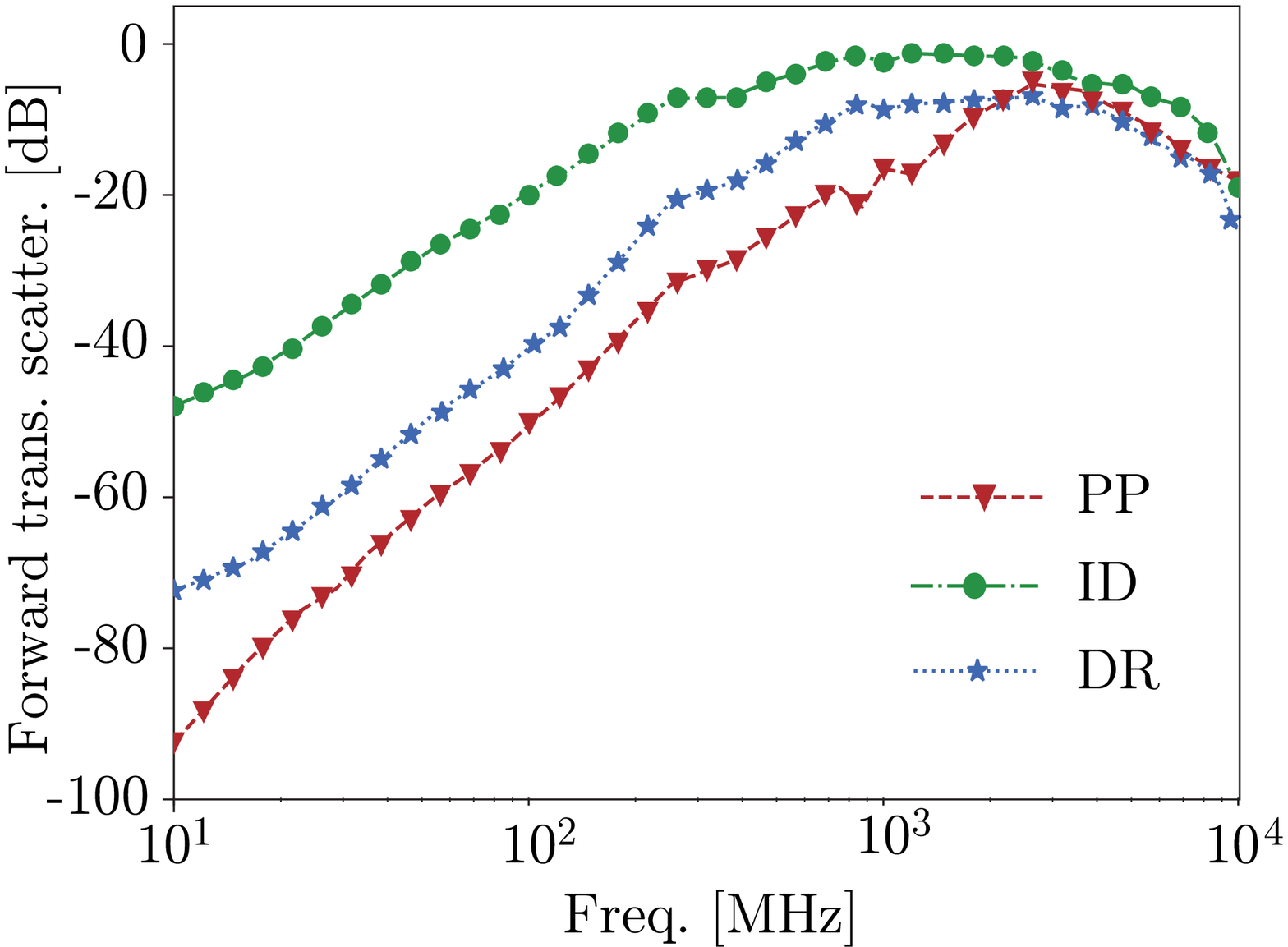}
         \caption{}
         \label{fig2:VNA}
     \end{subfigure}
     \hfill
    \begin{subfigure}[b]{0.85\columnwidth}
         \centering
         \includegraphics[width=\columnwidth]{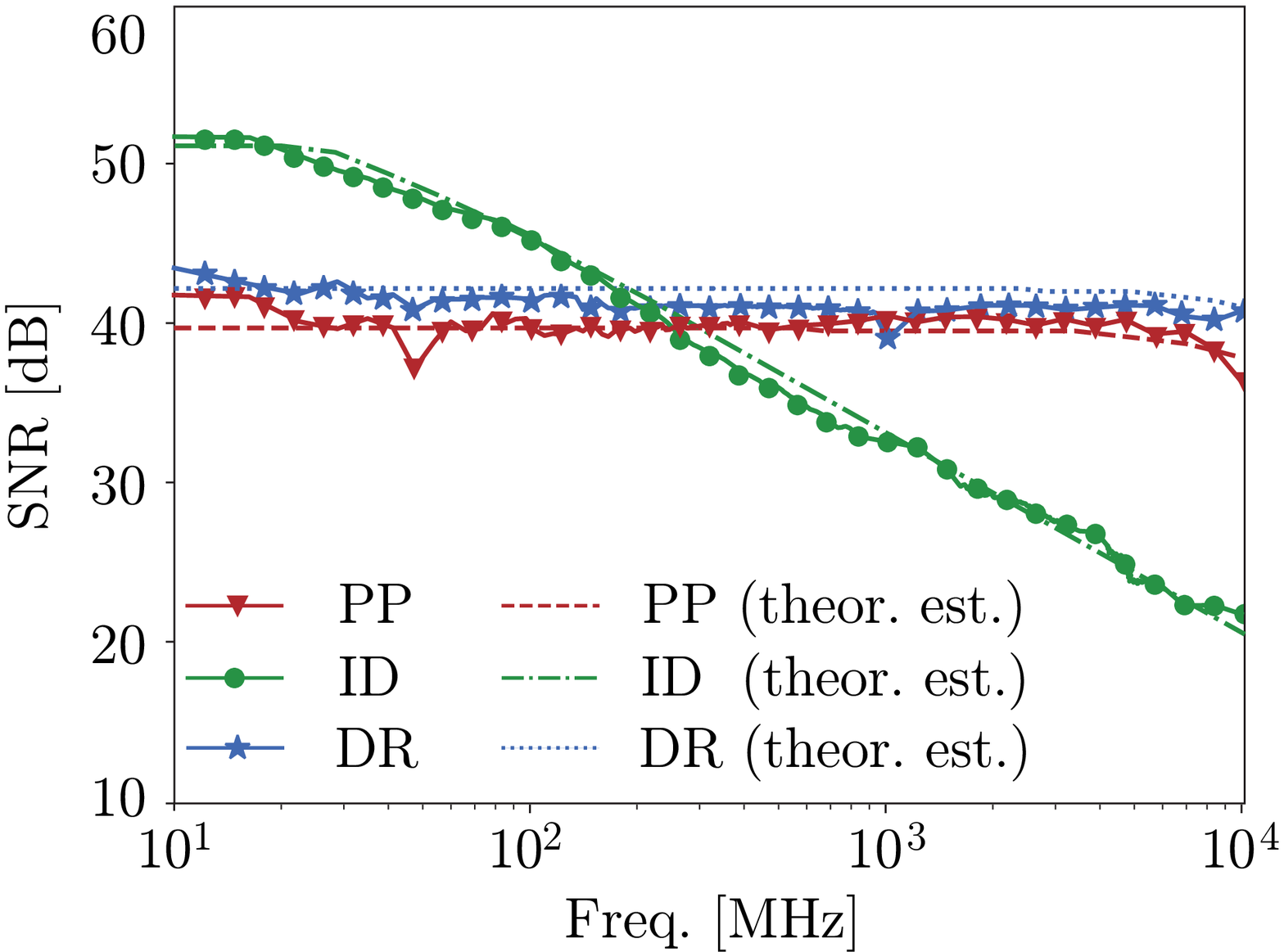}
         \caption{}
         \label{fig2:VNA_meas}
    \end{subfigure}
        \caption{Scattering parameters and SNR values. (a) Experimental forward transmission scattering $S_{21}$ parameters. (b) SNR from experimental results and theoretical estimations.}
        \label{fig:three graphs}
\end{figure}

The electrodes have been characterized at microwave frequencies using a ${0.01}$ to \SI{20}{\giga \hertz} two-port Vector Network Analyzer (VNA) with a \SI{91}{\deci \bel} of dynamic range and \SI{-94}{\deci \bel \m} sensitivity at the test port. The measured ${0.01}$ to \SI{10}{\giga \hertz} forward transmission scattering parameters $S_{21}$ for the three electrodes operating into the microchannel filled with water are reported in Fig.~\ref{fig2:VNA}.

Overall, it may be seen that the magnitude of the signals received by the three different coplanar electrodes increases with frequency, at a rate of around \SI{40}{\deci \bel/\si{decade}} (as expected, since the scattering power intensity scales as $f^4$ with frequency \cite{fan2015atom}). The responses having values that depend on the specific coupling for each geometry of electrodes where the PP electrode has the smallest $S_{21}$ measured magnitude, being roughly \SI{10}{\deci \bel} and \SI{30}{\deci \bel} smaller than the responses of the DR and ID electrodes, respectively.

Above \SI{1}{\giga \hertz} it has been observed, analyzing the reflection coefficient, that the PP electrode has a dominant capacitive effect (producing an equivalent short-circuit), while the ID and DR electrodes instead present an inductive dominant effect (producing an equivalent open-circuit). As a result, this dominant capacitive/inductive effect significantly reduces the sensing capabilities at large frequencies with the current design of electrodes, as it is experimentally observed and confirmed at \SI{10}{\giga \hertz} in Fig. \ref{fig3:10G}, having much lower detection levels. This capacitive/inductive effect above \SI{1}{\giga \hertz} results in an S-parameters dependence with frequency that is flat for the three electrodes, and consequently do not follow the previous scaling.

Sufficient level of detection is a necessary condition for sensing bioparticles, but it has to be combined with a measurable variation of the signal when a bioparticle is sensed. Hence, a more convenient parameter for characterizing the detection of bioparticles as they flow through the microchannel into the electromagnetic fields created by the electrodes is the signal-to-noise ratio (SNR); viz. the ratio between the signal, defined as the integrated contribution of the response magnitude of the peaks detected, over noise considered as the integrated baseline magnitude contribution over the area of the peaks' width.
In this regard, it is shown in Fig.~\ref{fig2:VNA_meas} the SNR levels measured for the three different electrodes compared to the theoretical estimation of the expected trends.
In the theoretical estimations, the trend is computed as the variation of the penetration depth of the hotspot, in which its dependence over frequency is evaluated from the analytical expressions derived in Section~\ref{sec:theory}.
In general, both PP and DR electrodes provide an almost flat response over the frequency range considered; this feature can be anticipated from the theoretical estimations. Instead, as it is also indicated by the theoretical estimations (decrease of its penetration depth), the ID electrode presents a constant decay as a function of frequency.

\subsection{Pulse levels detected}
As bioparticles flow through the electromagnetic fields of the electrodes, they produce signal pulses with SNR and time duration proportional to the produced electrical field disturbance, width of the electrodes and microfluidic flow rate, respectively, as summarized in Table~\ref{tab:pulse}.
\begin{figure}[ht!]
     \centering
     \begin{subfigure}[b]{0.75\columnwidth}
         \centering
         \includegraphics[width=\textwidth]{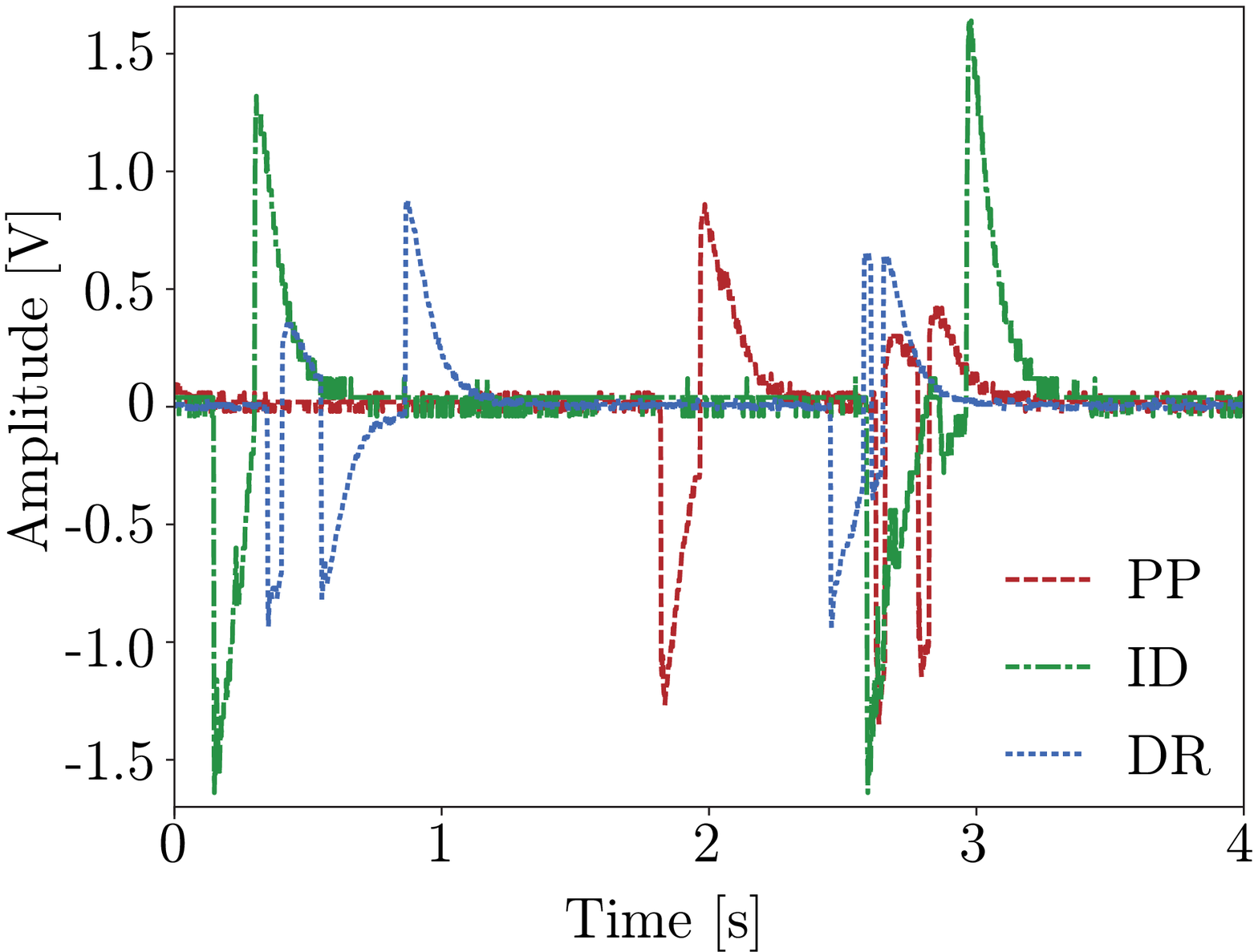}
         \caption{}
         \label{fig3:10M}
     \end{subfigure}
     \hfill
     \begin{subfigure}[b]{0.75\columnwidth}
         \centering
         \includegraphics[width=\textwidth]{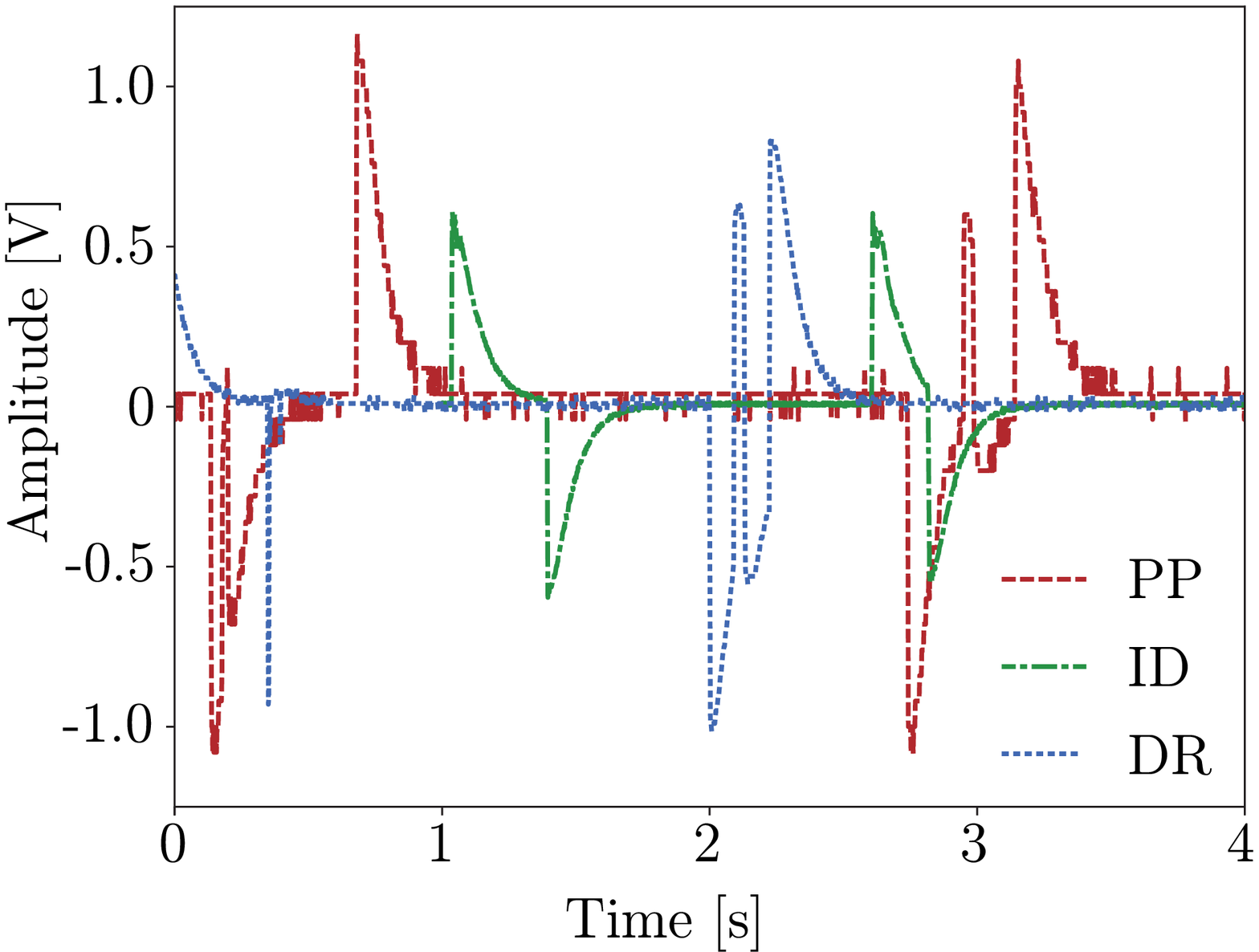}
         \caption{}
         \label{fig3:100M}
     \end{subfigure}
     \centering
     \begin{subfigure}[b]{0.75\columnwidth}
         \centering
         \includegraphics[width=\columnwidth]{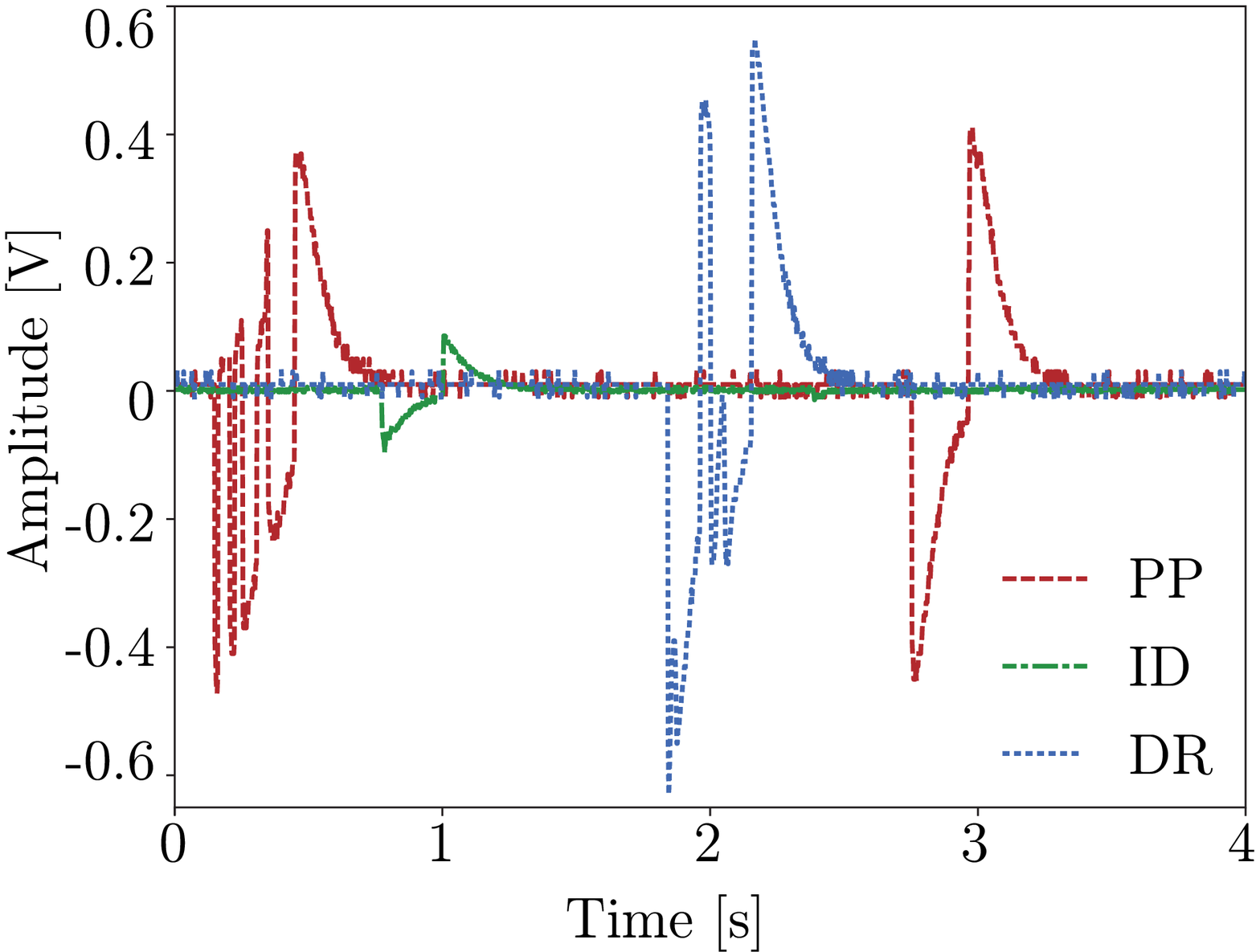}
         \caption{}
         \label{fig3:1G}
     \end{subfigure}
     \hfill
     \begin{subfigure}[b]{0.75\columnwidth}
         \centering
         \includegraphics[width=\columnwidth]{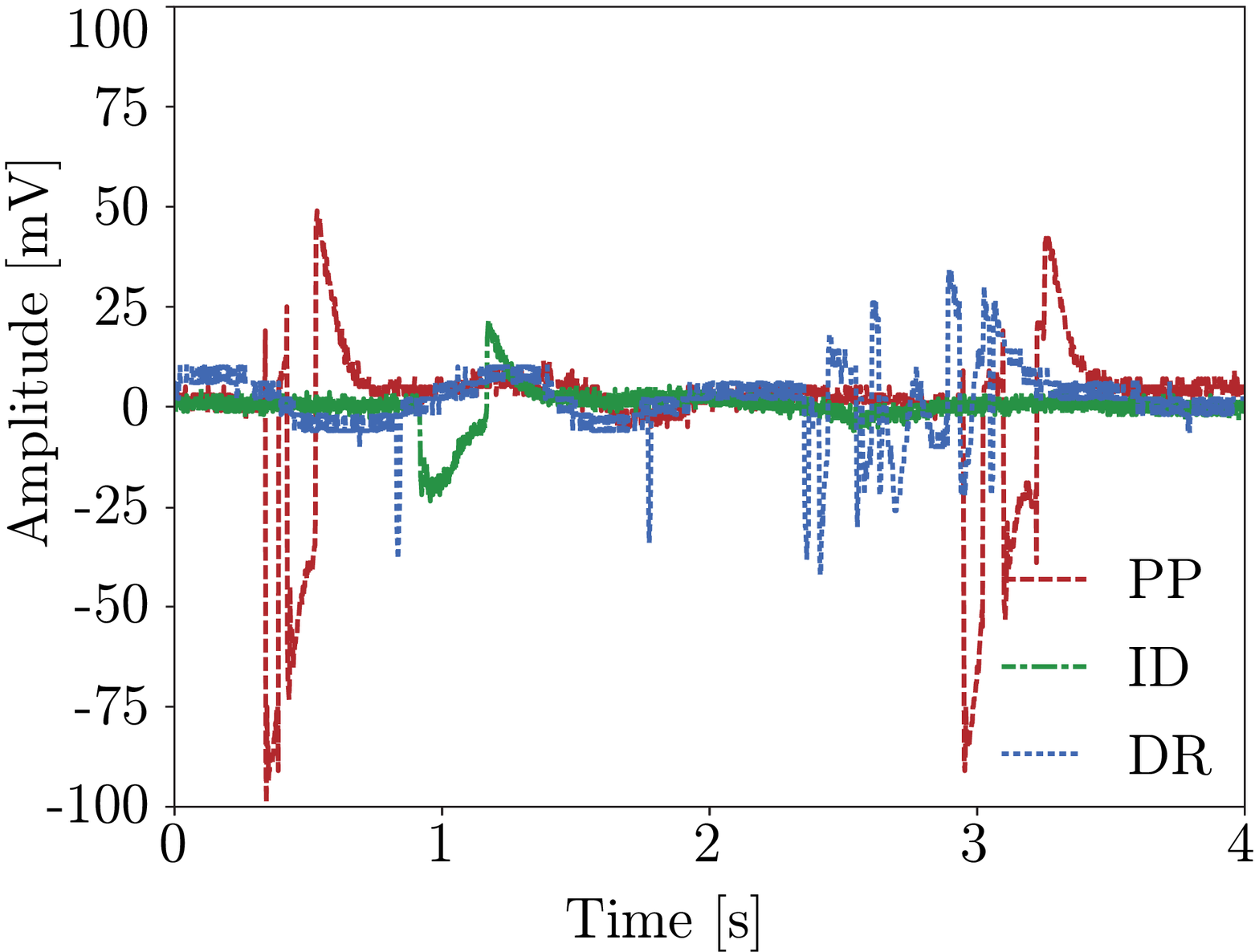}
         \caption{}
         \label{fig3:10G}
     \end{subfigure}
        \caption{Pulses detected from the bioparticles as a function of time at \SI{10}{\mega \hertz} (a), \SI{100}{\mega \hertz} (b), \SI{1}{\giga \hertz} (c) and \SI{10}{\giga \hertz} (d).}
        \label{fig:beads_measurements}
\end{figure}
\begin{table}[!ht]
\renewcommand{\arraystretch}{1.3}
\caption{Characteristics of measured pulses.}
\label{tab:pulse}
\centering
\begin{tabular}{|c|c|c|c|}
 \hline
 \multicolumn{1}{|c|}{\begin{tabular}[c]{@{}c@{}}Electrode \\ type \end{tabular}} &  \multicolumn{1}{c|}{\begin{tabular}[c]{@{}c@{}}SNR freq. dB\\ variation {[}\%{]}\end{tabular}} & \multicolumn{1}{c|}{\begin{tabular}[c]{@{}c@{}}Pulse \\ width {[}ms{]}\end{tabular}} & \multicolumn{1}{c|}{\begin{tabular}[c]{@{}c@{}}Interpulse \\ interval {[}s{]}\end{tabular}} \\
 \hline
PP & 14.01 & 516 & 1.05 \\
ID & 26.84 & 618 & 1.33 \\
DR & 11.93 & 809 & 1.71 \\
 \hline
\end{tabular}
\end{table}

Figure~\ref{fig:beads_measurements} shows the AC-coupled signal received by the system for \SI{10}{\mega \hertz}, \SI{100}{\mega \hertz}, \SI{1}{\giga \hertz}, and \SI{10}{\giga \hertz}, when a \SI{250}{\milli \volt} input signal is fed to each of the three different electrodes considered. As anticipated in Fig.~\ref{fig2:VNA_meas}, the ID electrode behaves better at \SI{10}{\mega \hertz}, while PP and DR electrodes perform better at higher frequencies. In this sense, the signal received from the ID electrode has an amplitude of approximately \SI{1.5}{\volt} at the lowest frequency, while the amplitudes for the PP and DR electrodes are roughly \SI{0.8}{\volt}. For the ID electrodes, when operating at \SI{100}{\mega \hertz}, the amplitude decays rapidly to \SI{0.5}{\volt}, which is below the values for PP and DR at approximately \SI{0.8}{\volt}.
Moreover, at \SI{1}{\giga \hertz}, the amplitude of the PP and DR pulses decrease roughly to \SI{0.3}{\volt}, while the amplitude for ID is below \SI{0.2}{\volt}. Finally, it is important to note that, at \SI{10}{\giga \hertz}, the capacitive and inductive effects described above significantly reduce the detection performance of the three electrodes studied; viz. the amplitudes are reduced to pulses below \SI{50}{\milli \volt}. Even so, the responses of the bioparticles remain within detectable levels in this study.

\section{Frequency dependence discussion}      \label{sec:discussion}
The combination of a superheterodyne architecture (with broadband coverage) and a lock-in detection receiver (capable of achieving very high sensitivities), when compared to a homodyne detection, presents a reduction of \SI{3}{\deci \bel} in its SNR [noise increases at the intermediate frequency (IF) detection stage as a result of the two image components; right and left sides of the central frequency], but provides the significant advantage of being able to operate up to very high microwave frequencies only limited by the parameters of the transmission line, electrode geometries, and the operating frequency range of the spectrum analyzer.
In this regard, the system proposed presents significant advantages when it comes to characterizing bioparticles down to eventually millimeter wave frequencies, since it allows to reach frequencies higher than the common operational frequency limit for lock-in amplifiers (\SI{500}{\mega \hertz}).

In this work, the pairing of a grounded coplanar waveguide (GCPW), acting as a 10 GHz resonance free transmission line, with PP and DR broadband coplanar electrodes allowed to reach frequencies up to \SI{10}{\giga \hertz}.
Therefore, a solution combining PP (capacitive reactance) and DR (inductive reactance) electrodes will allow to reduce the limiting reactance effect, which, together with a GCPW feeding line, will provide a significant increase of the operational frequency.

In general, the SNR obtained for the PP and DR electrodes is mostly uniform at approximately \SI{45}{\deci \bel} for the range of frequencies measured. Moreover, the ID electrodes provide SNR of \SI{50}{\deci \bel} at low frequencies, which decrease to roughly \SI{20}{\deci \bel} at high frequencies.
Nonetheless, when considering the expected \SI{30}{\deci \bel} signal decrease due to a $1000\times$ volume reduction of bioparticles, the measurements obtained indicate that detection levels of \SI{1}{\micro \meter}-sized microorganisms (e.g., bacteria) fall within the experimental demonstrated capabilities of the system.

\subsection{Extension to higher frequency ranges}

In this work, the frequency detection range of bioparticles has been extended into the \si{\giga \hertz} range (up to \SI{10}{\giga \hertz}) allowing to enter into the $\gamma-$dispersion region to enable the identification of cell types~\cite{pohl1966separation,schwan1957electrical}, by making use of a superheterodyne receiver, in contrast to common lock-in amplifier systems which only reach hundreds of \si{\mega \hertz}. To compare the detection systems commonly used at \si{\mega \hertz} frequencies with the system described in this work, capable of working at \si{\giga \hertz}, it is shown in Table~\ref{tab:FullRange_measurement} the detection results of \SI{10}{\micro \meter} bioparticles consisting of polystyrene beads with both systems for different frequency ranges. In particular, it is shown the measured peaks with the three different coplanar electrodes in the range \SI{100}{\kilo \hertz} to \SI{10}{\mega \hertz} for the lock-in amplifier, and the range \SI{10}{\mega \hertz} to \SI{10}{\giga \hertz} for the superheterodyne receiver.

\begin{table}[!ht]
\caption{Measured peaks of \SI{10}{\micro\meter} beads with the different electrodes in the range \SI{100}{\kilo\hertz} to \SI{10}{\giga\hertz} using a lock-in amplifier based system and the superheterodyne receiver.}
\label{tab:FullRange_measurement}
\resizebox{\columnwidth}{!}{%
\begin{tabular}{l|cc|cc|cc|}
\cline{2-7}
\multicolumn{1}{c|}{} & \multicolumn{2}{c|}{\begin{tabular}[c]{@{}c@{}}PP\\ Peak values {[}V{]}\end{tabular}} & \multicolumn{2}{c|}{\begin{tabular}[c]{@{}c@{}}ID\\ Peak values {[}V{]}\end{tabular}} & \multicolumn{2}{c|}{\begin{tabular}[c]{@{}c@{}}DR\\ Peak values {[}V{]}\end{tabular}} \\ \hline
\multicolumn{1}{|c|}{\begin{tabular}[c]{@{}c@{}}Freq. \\ $[$MHz$]$\end{tabular}} & \multicolumn{1}{c|}{LIA} & Superhet. & \multicolumn{1}{c|}{LIA} & Superhet. & \multicolumn{1}{c|}{LIA} & Superhet. \\ \hline
\multicolumn{1}{|l|}{$10^{-1}$} & \multicolumn{1}{c|}{1.18} & $-$ & \multicolumn{1}{c|}{1.75} & $-$ & \multicolumn{1}{c|}{0.88} &  $-$\\ \hline
\multicolumn{1}{|l|}{$10^{0}$} & \multicolumn{1}{c|}{1.24} & $-$ & \multicolumn{1}{c|}{1.23} & $-$ & \multicolumn{1}{c|}{0.93} & $-$ \\ \hline
\multicolumn{1}{|l|}{$10^{1}$} & \multicolumn{1}{c|}{1.16} & 0.90 & \multicolumn{1}{c|}{0.95} & 1.43 & \multicolumn{1}{c|}{0.86} & 0.71 \\ \hline
\multicolumn{1}{|l|}{$10^{2}$} & \multicolumn{1}{c|}{$-$} & 1.22 & \multicolumn{1}{c|}{$-$} & 0.58 & \multicolumn{1}{c|}{$-$} & 0.81 \\ \hline
\multicolumn{1}{|l|}{$10^{3}$} & \multicolumn{1}{c|}{$-$} & 0.39 & \multicolumn{1}{c|}{$-$} & 0.09 & \multicolumn{1}{c|}{$-$} & 0.51 \\ \hline
\multicolumn{1}{|l|}{$10^{4}$} & \multicolumn{1}{c|}{$-$} & 0.09 & \multicolumn{1}{c|}{$-$} & 0.02 & \multicolumn{1}{c|}{$-$} & 0.03 \\ \hline
\end{tabular}
}
\end{table}
The peak values for the different coplanar electrodes as a function of frequency follow a continuity trend when passing from the lock-in amplifier detection technique to the superheterodyne-based detection systems, which overlap at \SI{10}{\mega \hertz}. The values follow a decreasing trend with frequency, which is consistent with the decreasing response of the electrodes with frequency as explained in the previous subsection.

Starting from the standard definition of a bioparticle as “a particle of biological material”, in this work the concept is extended to also include beads of size $1$ to \SI{10}{\micro \meter} (similar to microorganisms) with electric parameters capable to electromagnetically mimic bioparticles in the frequency range \SI{0.01}{\mega \hertz} to \SI{10}{\giga \hertz}, and with an especial interest on the upper microwave region. Typically, polystyrene beads are utilized for biosensor assessment and calibration as it is a reliable process for adjusting instruments to produce an accurate physical measurement~\cite{ormerod2000flow,givan2013flow}. Therefore, the use of polystyrene beads (i) helps ensure that instruments operate correctly, (ii) facilitates the reproducibility of experimental results, and (iii) provides accuracy for wider parameter studies~\cite{wang2017standardization}. In addition, based on the analysis of signal detection capabilities above, it has been estimated the similarities, differences and limitations for a selected group of common bioparticles (Yeast cells and E. Coli bacteria) and compared to \SI{1}{\micro \meter} and \SI{10}{\micro \meter} polystyrene beads. 

\begin{table}[!ht]
\caption{Typical size range and dielectric complex relative permittivity $\varepsilon^*_r=\varepsilon'_r-j\varepsilon''_r$ for a polystyrene bead, Yeast cell and E. Coli at different frequencies.}
\label{tab:permittivities}
\centering
\resizebox{\columnwidth}{!}{%
\begin{tabular}{c|ccc|}
\cline{2-4}
 & \multicolumn{3}{c|}{\begin{tabular}[c]{@{}c@{}}Bioparticle \end{tabular}} \\ \hline
\multicolumn{1}{|c|}{\begin{tabular}[c]{@{}c@{}}Freq.\\ $[$\si{\mega \hertz}$]$\end{tabular}} & \multicolumn{1}{c|}{\begin{tabular}[c]{@{}c@{}}Polystyrene bead\\ Size (1-10 \si{\micro \meter})\end{tabular}} & \multicolumn{1}{c|}{\begin{tabular}[c]{@{}c@{}}Yeast cell\\ Size (1-6 \si{\micro \meter})\end{tabular}} & \begin{tabular}[c]{@{}c@{}}E. Coli\\ Size (1-2 \si{\micro \meter})\end{tabular} \\ \hline
\multicolumn{1}{|c|}{$10^{-1}$} & \multicolumn{1}{c|}{$3.17-j0.03$} & \multicolumn{1}{c|}{$500-j34168$} & $~~225-j59345$ \\ \hline
\multicolumn{1}{|c|}{$10^0$} & \multicolumn{1}{c|}{$3.12-j0.04$} & \multicolumn{1}{c|}{$300-j3776~$} & $~~110-j6294~$ \\ \hline
\multicolumn{1}{|c|}{$10^1$} & \multicolumn{1}{c|}{$3.07-j0.04$} & \multicolumn{1}{c|}{$110-j413~~$} & $75.38-j659~~$ \\ \hline
\multicolumn{1}{|c|}{$10^2$} & \multicolumn{1}{c|}{$2.94-j0.04$} & \multicolumn{1}{c|}{$~90-j46.75$} & $75.26-j59.52$ \\ \hline
\multicolumn{1}{|c|}{$10^3$} & \multicolumn{1}{c|}{$2.87-j0.03$} & \multicolumn{1}{c|}{$~76-j5.39~$} & $75.05-j6.02~~$ \\ \hline
\multicolumn{1}{|c|}{$10^4$} & \multicolumn{1}{c|}{$2.82-j0.02$} & \multicolumn{1}{c|}{$~63-j0.64~$} & $~74.8-j0.61~$ \\ \hline
\end{tabular}%
}
\end{table}

The similarities between polystyrene beads and microorganisms can be accommodated in size by choosing beads of similar diameter as Yeast cells and/or E. Coli bacteria. The values of the dielectric complex relative permittivity $\varepsilon^{*}_{r,p}$ of these bioparticles are obtained from~\cite{von1954dielectric,asami1976dielectric,russel2018high,bai2006dielectric} and are reported in Table~\ref{tab:permittivities}.

\begin{figure}[ht!]
    \centering
    \includegraphics[width=0.85\columnwidth]{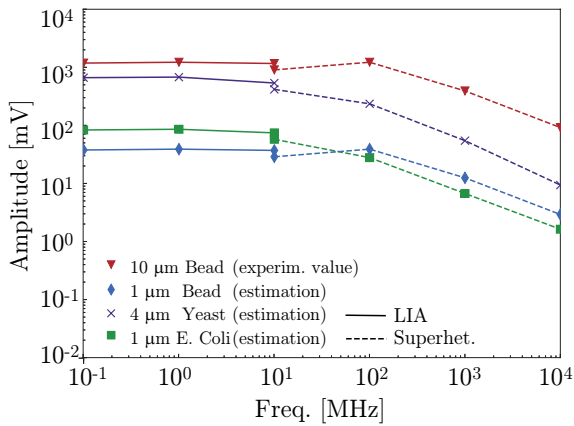}
    \caption{Estimation, based on experiments with \SI{10}{\micro\meter} polystyrene beads, of the absolute voltage detected for different bioparticles in the frequency range \SI{100}{\kilo\hertz} to \SI{10}{\giga\hertz}.}
    \label{fig:BioParticlesTheoreticalEstimation}
\end{figure}

When detecting bioparticles with electromagnetic sensors, the two most significant parameters are the diameter of the bioparticle being sensed and the Clausius-Mossotti factor, which is related to the dielectric complex relative permittivity of the bioparticle and medium.
Particularly, as shown in Fig.~\ref{fig:BioParticlesTheoreticalEstimation}, the detection signal voltage scales with the particle’s diameter as a power of $3/2$ and as a power of $1/2$ with the permittivity of the biomaterial.

Experimental detection values of \SI{10}{\micro\meter} polystyrene beads from Table~\ref{tab:FullRange_measurement} have been used as a reference to estimate the results in Fig.~\ref{fig:BioParticlesTheoreticalEstimation}. As a first-order approximation, the expected detection voltage values have been calculated using PP electrodes for (i) a \SI{4}{\micro\meter} Yeast cell and E. Coli, and (ii) a \SI{1}{\micro\meter} polystyrene bead, and the three of them compared to the \SI{10}{\micro\meter} polystyrene bead. These calculations utilize the mentioned sizes and dielectric complex relative permittivity provided in Table~\ref{tab:permittivities}. The results for the detection levels confirm the larger importance of size with respect to dielectric complex relative permittivity contrast, and, thus, support the use of polystyrene beads to analyze the sensing of bioparticles. They provide also an indication of the eventual convenience of jointly tuning the geometries of the electrode and microchannel to optimize the response of the system. Moreover, the availability of the frequency response when correlated with its frequency permittivity trend (Table~\ref{tab:permittivities}) may enable extending the use of the system for bioparticle differentiation.

\section{Conclusions}     \label{sec:conclusions}
Advancing the scientifico-technical frontier of sensing microorganisms provides many opportunities to generate new discoveries and develop novel applications in a wide range of areas, especially in biotechnology.
In this regard, the multidisciplinary system presented and experimentally demonstrated in this work to sense bioparticles is based on combining microwave communication techniques, which is a virtually unexplored frequency range window for interacting with microorganisms, with microfluidics for an efficient hydrodynamic focusing of single-particle flows.
The resulting microfluidics-based superheterodyne microwave detection system is demonstrated to be significantly adequate for operating and sensing micron-sized elements, like for example bacteria and cells.

Generally, coplanar-type electrodes are good solutions for diverse electromagnetic applications, as well as for detecting bioparticles in microfluidic platforms, especially when parallel facing electrodes cannot be mounted.
In particular, the analytical expressions of the spatial field distributions for the three electrode configurations considered have allowed to carefully infer insight and perform validation of the results obtained from experiments.
Moreover, the theoretical expressions developed and presented warrant the possibility of designing these type of electrodes for other applications targeted at higher frequency ranges and/or for other system arrangements.

Additionally, the novel superheterodyne microwave receiver proposed enables high-throughput of bioparticle detection for a wide and robust range of microwave frequencies with optimized SNR capabilities.
In particular, the receiver architecture proposed facilitates operating at microwave frequencies, resulting in a novel approach since, up to present, the solutions presented in the literature were limited to (complex) very specific and/or individual-frequency-operation systems.

As future work, in the mid- to long-term, the approach proposed and technology demonstrated will be further optimized for applications related to robust detection of bioparticles for in-field operation by focusing on increasing the (i) maximum operational frequency, and (ii) detection sensitivity.



\section*{Acknowledgments}
This work was financially supported by CICYT PID2019-107885GB-C31, MDM2016-0600, and the Formaci\'on de Personal Investigador (PID2019-107885GB-C31, PRE2020-093895) and Beatriz Galindo (BGP18$/$00026) programs of the Ministerio de Ciencia, Innovaci\'on y Universidades (Spain).


\bibliographystyle{IEEEtran}
\bibliography{references}

%

\vskip -1\baselineskip plus -1fil

\begin{IEEEbiography}[{\includegraphics[width=1in,height=1.25in,clip,keepaspectratio]{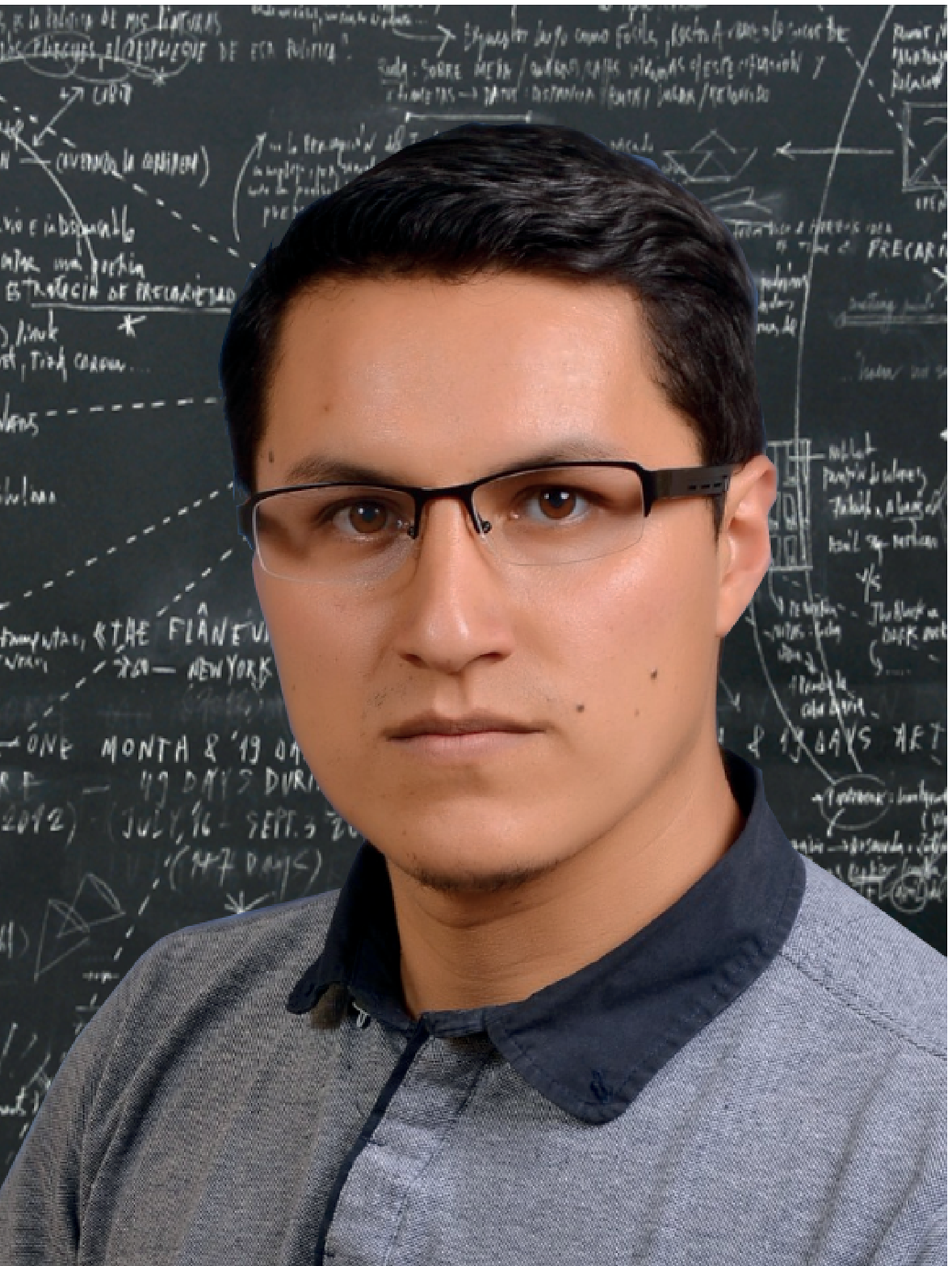}}]{César Palacios}, He received the B.S. degree in Telecommunication and Electronics Engineering in 2013 from the Private Technical University of Loja UTPL (Ecuador), and the M.S. degree in Electronics Engineering from the University of Calabria (Italy) in 2017. He is currently pursuing the Ph.D. degree in the Signal Theory and Communications (TSC) Department, within the research group of microwave interaction with living organisms, CommSensLab, UPC. He has held positions at ALCATEL-LUCENT (2013-15), Corporaci\'on Nacional de Telecomunicaciones-CNT (2015-15), external researcher at National University of Chimborazo (2018-), and research support technician at the Signal Theory and Communications (TSC) Department, UPC (2020-21). He is currently working on micro-systems design and manufacturing for communication with living organisms and sensing at X-wave frequencies.
\end{IEEEbiography}
\vskip -1\baselineskip plus -1fil

\begin{IEEEbiography}[{\includegraphics[width=1in,height=1.25in,clip,keepaspectratio]{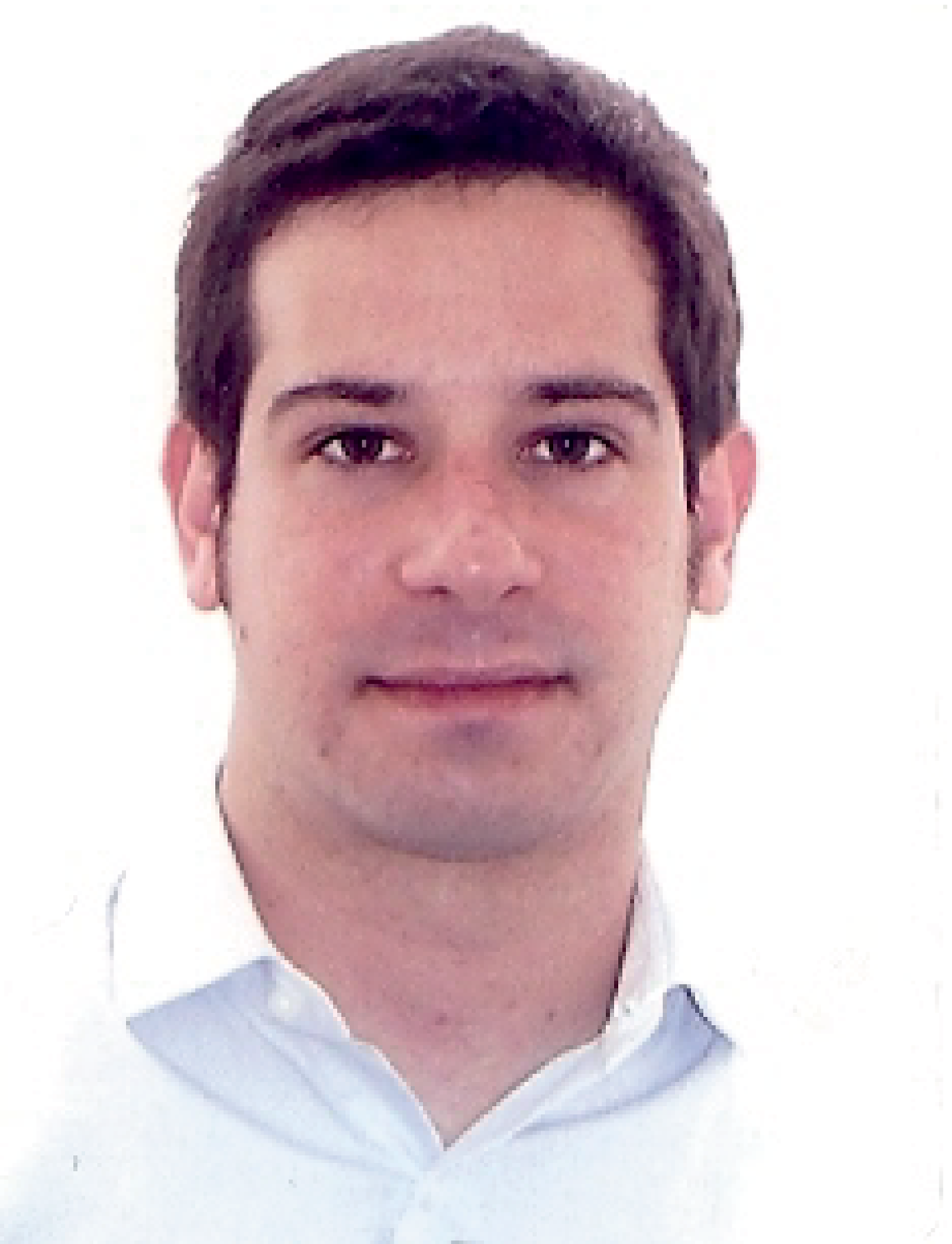}}]{Marc Jofre}
Ph.D., He obtained his Electrical Engineering (Telecom) Degree in 2008 from the Technical University of Catalonia (UPC) - BarcelonaTech (Spain), jointly with the Technical University of Delft (The Netherlands). In 2013, he obtained a Ph.D. in Photonic Sciences from ICFO - The Institute of Photonic Sciences (Spain). From 2018, he has held different positions at Technical University of Catalonia (UPC), FPHAG – Fundaci\'o Privada Hospital Asil de Granollers, and Max-Planck Institute for Quantum Optics. He has 16 peer-reviewed papers, contributed to more than 20 conferences, 8 granted patents/patent applications, and participated in several funded projects (national and international; competitive and non-competitive). He has extensive experience in innovation and research in managing projects, intellectual property, business, and technology exploitation for quantum communication, sensing systems, nonlinear material physics characterization, biophysical detectors and platforms for quantifying microorganisms, and managing health technology projects.
\end{IEEEbiography}
\vskip -1\baselineskip plus -1fil

\begin{IEEEbiography}[{\includegraphics[width=1in,height=1.25in,clip,keepaspectratio]{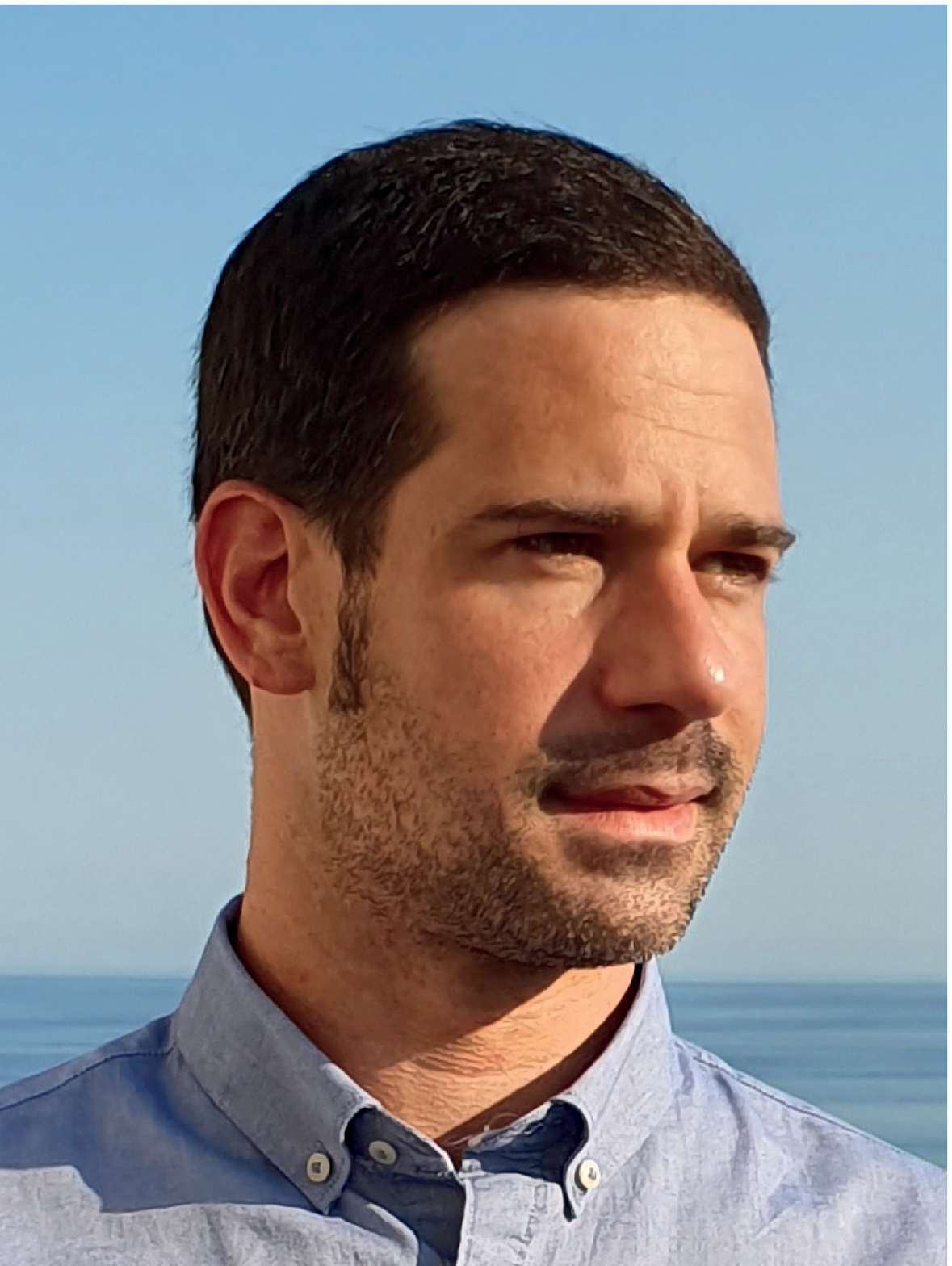}}]{Llu\'is Jofre}
Ph.D., Beatriz Galindo Professor in the Department of Fluid Mechanics at the Technical University of Catalonia (UPC) - BarcelonaTech (Spain).
He obtained his Mechanical Engineering Degree in 2008 from the Technical University of Catalonia - BarcelonaTech (Spain), jointly with the KTH - Royal Institute of Technology (Sweden).
In 2014, he obtained a Ph.D. in Fluid Mechanics and Thermal Engineering from the Technical University
of Catalonia - BarcelonaTech (Spain).
From 2015 to 2020 he was a postdoctoral researcher at the Center for Turbulence Research at Stanford University (USA).
His main research areas include microfluidics, biophysical fluids, high-pressure supercritical flows, two-phase flows, modeling and computational studies of turbulence in multiphysics environments, uncertainty quantification, and data science in fluid mechanics.
He has contributed to 25 peer-reviewed papers and more than 30 conferences, and participated in 10 funded competitive/non-competitive projects nationally and internationally.
\end{IEEEbiography}
\vskip -1\baselineskip plus -1fil

\begin{IEEEbiography}[{\includegraphics[width=1in,height=1.25in,clip,keepaspectratio]{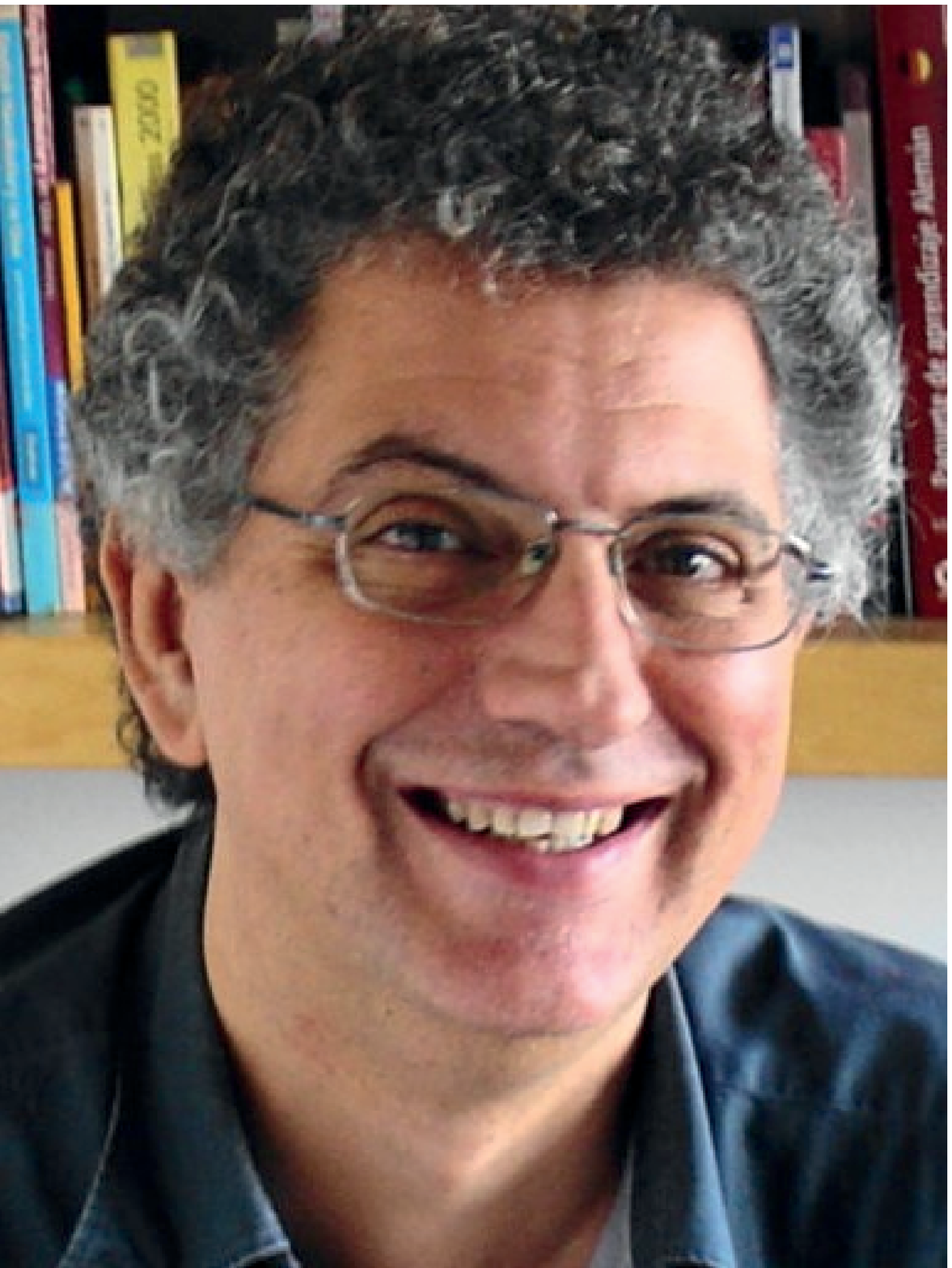}}]{Jordi Romeu} Ph.D., (IEEE Fellow) He received the Ingeniero de Telecomunicación and Doctor Ingeniero de Telecomunicación degrees from the Technical University of Catalonia (UPC) - BarcelonaTech (Spain) in 1986 and 1991, respectively. In 1985, he joined the AntennaLab, Signal Theory and Communications Department (UPC), where he is currently a Full Professor involved in antennas near-field measurements, electromagnetic scattering and imaging, and system miniaturization for wireless and sensing industrial and bio applications. In 1999, he was a Visiting Scholar at the Antenna Laboratory of the University of California at Los Angeles (USA) on a NATO Scientific Program Scholarship, and at the University of California at Irvine (USA) in 2004. He holds several patents, and has published 60 refereed articles in international journals and 80 conference proceedings. He was a Grand Winner of the European IT Prize, awarded by the European Commission for his contributions in the development of fractal antennas in 1998. He has been involved in the creation of several spin-off companies.
\end{IEEEbiography}
\vskip -1\baselineskip plus -1fil

\begin{IEEEbiography}[{\includegraphics[width=1in,height=1.25in,clip,keepaspectratio]{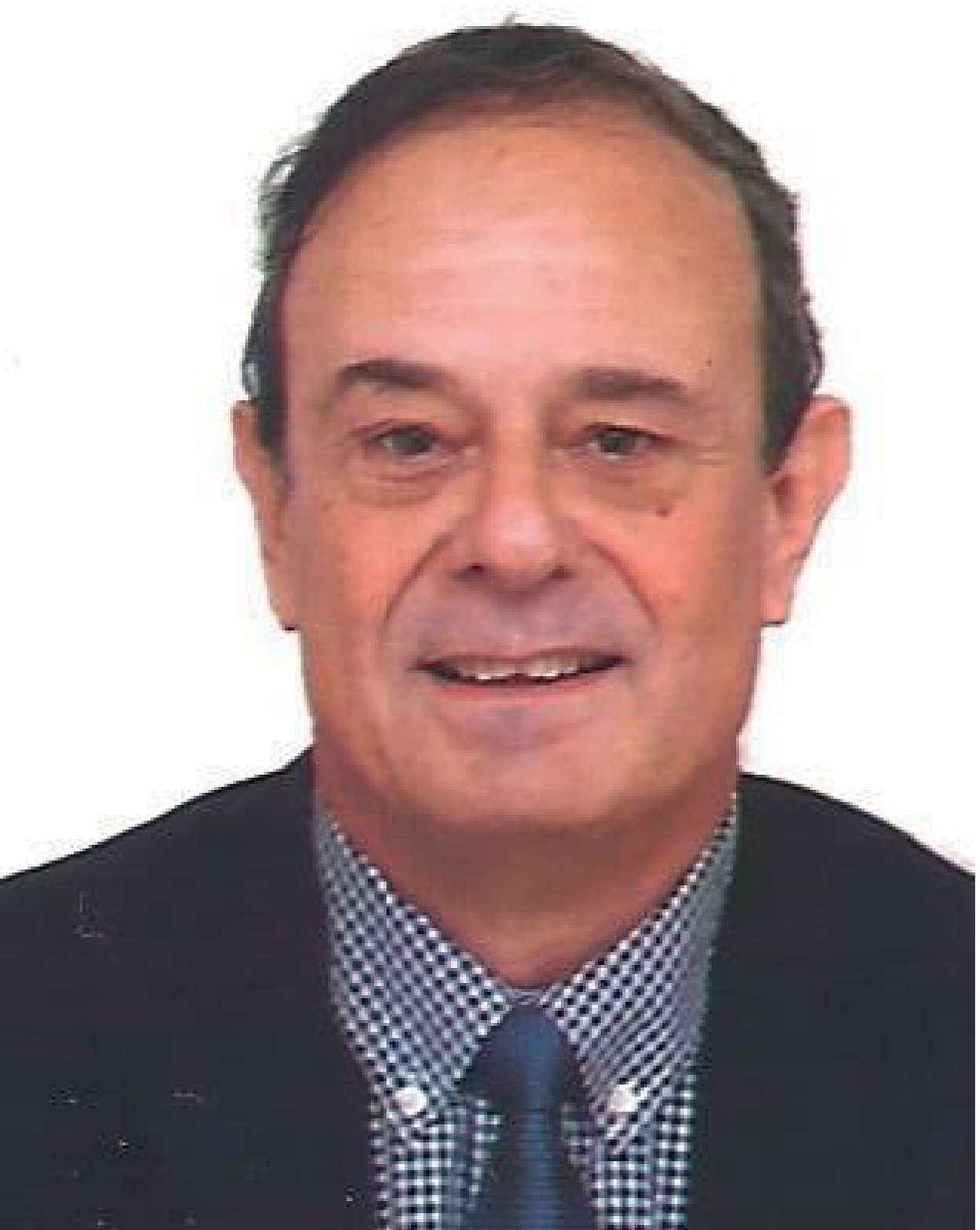}}]{Luis Jofre-Roca} Ph.D., (IEEE Fellow, 2010) He received the M.Sc. (Ing.) and Ph.D. (Doctor Eng.) degrees in Electronic Engineering (Telecommunication Engineering) from the Technical University of Catalonia (UPC), Barcelona, Spain, in 1978 and 1982, respectively. He has been visiting professor at the École Supérieure d’Electricité Paris (1981-82), with the Georgia Institute of Technology, Atlanta (Fulbright Scholar, 1986-87), and at the University of California, Irvine, CA (2001-02). Director (1989-94) of the Telecommunication Engineering School, UPC, Vicepresident of the UPC (1994-2000), and General Director and Secretary for Catalan Universities and Research (2011-2016). Director of the Catalan Research Foundation (2002-04), Director of the UPC-Telefonica Chair on Information Society Future Trends (2003-), Principal Investigator of the 2008-13 Spanish Terahertz Sensing Lab Consolider Project, Director of the UPC-SEAT Chair on the Future of Automotive, Research Leader of the 2017–2020 CommSensLab Maria de Maeztu Project, Academic Director of the European Consortium for Future Urban Mobility (Carnet) and Chairman of the EIT-Urban Mobility European Association. He has authored more than 200 scientific and technical papers, reports, and chapters in specialized volumes. Research interests include antennas, electromagnetic scattering and imaging, system miniaturization for wireless and sensing for industrial, scientific and medical applications. Current work focuses on Connected Reconfigurable Autonomous Vehicles for Urban Mobility, Massive MIMO Antennas and Microorganism wireless interaction.
\end{IEEEbiography}


\end{document}